\documentclass[11pt,a4paper]{article}

\usepackage[T1]{fontenc}
\usepackage{lmodern}
\usepackage{amsmath,amssymb,amsthm}
\usepackage[margin=1in]{geometry}
\usepackage{microtype}
\usepackage{array,booktabs,tabularx,longtable}
\usepackage{tikz}
\usetikzlibrary{matrix}
\usepackage{needspace}
\usepackage[colorlinks=true,linkcolor=blue,citecolor=blue,urlcolor=black]{hyperref}

\numberwithin{equation}{section}

\newcommand{\F}{\mathbb{F}}
\newcommand{\DDT}{\operatorname{DDT}}
\newcommand{\BCT}{\operatorname{BCT}}
\newcommand{\EA}{\operatorname{EA}}
\newcommand{\CCZ}{\operatorname{CCZ}}

\newtheorem{theorem}{Theorem}[section]
\newtheorem{corollary}[theorem]{Corollary}
\newtheorem{lemma}[theorem]{Lemma}
\newtheorem{proposition}[theorem]{Proposition}
\newtheoremstyle{boldremark}
  {.5\topsep}{.5\topsep}
  {\normalfont}{}
  {\bfseries}{.}{.5em}
  {\thmname{#1}\thmnumber{ #2}\thmnote{ (#3)}}
\theoremstyle{boldremark}
\newtheorem{remark}[theorem]{Remark}

\title{New Classes of CCZ-Inequivalent APN Functions\\
 with Boomerang Uniformity One in Odd Characteristic:\\
 Differential and Boomerang Spectra}
\author{%
Namhun Koo\textsuperscript{1}, Soonhak Kwon\textsuperscript{2}\\[0.35em]
\small Email: \href{mailto:komaton@skku.edu}{komaton@skku.edu},
\href{mailto:shkwon@skku.edu}{shkwon@skku.edu}\\[0.35em]
\footnotesize\textsuperscript{1}Institute of Basic Science, Sungkyunkwan University, Suwon, Korea\\
\footnotesize\textsuperscript{2}Department of Mathematics, Sungkyunkwan University, Suwon, Korea
}
\date{\small November 07, 2026}

\begin{document}
\maketitle
\vspace{-2.5em}

\begin{abstract}
Let $q=p^n>3$ be an odd prime power and let $g:\F_q\to\F_q$ be an
even perfect nonlinear (PN) function with $g(0)=0$.
We construct almost perfect nonlinear (APN) functions
$G_{\lambda,M}(x)=g(x)+M(x)+\lambda(x^{q-1}-1)$,
where $\lambda\in\F_q^*$, $-\lambda\notin g(\F_q)$, and $M$ is an
$\F_p$-linearized permutation polynomial.
These one-point modifications of $g+M$ have boomerang uniformity one,
the smallest possible value for APN functions in odd characteristic.
We give complete differential and boomerang spectrum formulas and obtain
closed-form evaluations for selected constructions based on Coulter--Matthews
monomials and Dembowski--Ostrom polynomials.
For $q>7$, comparison of differential spectra establishes CCZ-inequivalence to all
APN monomial families listed in the paper and every APN binomial in
the original Ness--Helleseth family or its extension by Zeng et al.
We characterize CCZ-equivalence within our construction, showing that
any two linearly inequivalent even PN functions over the same field yield
CCZ-inequivalent APN functions.
We obtain three pairwise CCZ-inequivalent APN functions with boomerang
uniformity one over $\F_{p^{15\ell}}$ for every $\ell\ge1$, six over
$\F_{p^{120r}}$ for every $r\ge1$, and unbounded numbers of such classes
in characteristic three.
We further prove that CCZ-equivalence and EA-equivalence coincide
for functions at Hamming distance at most $m$ from a PN function,
provided that $m(m+1)<q-q/p$.  We also show that all our APN
functions have the maximal possible algebraic degree $n(p-1)$.
Finally, we exhibit both CCZ-equivalent and CCZ-inequivalent APN pairs
with boomerang uniformity one at Hamming distance one.
\end{abstract}

\section{Introduction}\label{sec:introduction}

Let $q=p^n$ be an odd prime power.  Differential and boomerang
properties are central criteria in the study of cryptographic functions
over a finite field $\F_q$.  Differential uniformity was introduced by
Nyberg~\cite{Nyberg94}; functions with differential uniformity one or
two are called perfect nonlinear (PN) or almost perfect nonlinear (APN),
respectively.  The boomerang connectivity table, introduced by Cid et
al.~\cite{CidEtAl18} for permutations, admits an inverse-free formulation
that extends to arbitrary functions~\cite{LiQuSunLi2019}.
We write $\delta_F$ and $\beta_F$ for the differential and boomerang
uniformities of a function $F:\F_q\to\F_q$.
The differential and boomerang spectra record the number of occurrences
of each value in the corresponding tables and provide more detailed
information than the uniformities alone.  The precise definitions and conventions are
given in Section~\ref{sec:preliminaries}.

Recent work in odd characteristic has produced functions with the very
small boomerang uniformities $0,1,2$.  Every PN function has boomerang
uniformity zero, but non-PN functions can also have this property.
Several examples arise from Ness--Helleseth-type binomials
$H_{r,u}(x)=x^r(1+u\chi(x))$, where $\chi$ is the quadratic character
of $\F_q$, extended by $\chi(0)=0$.
Lyu et al.~\cite{LyuWangZheng2024} determined the differential
and boomerang spectra for $r=q-2$ and $u=\pm1$ when
$q\equiv3\pmod4$, obtaining classes
with boomerang uniformity zero or one.
Koo and Kwon~\cite{KooKwon2026} studied $r=(q+1)/4$ and $u=\pm1$
for $q\equiv3\pmod4$ and obtained classes with boomerang uniformity
zero, one, or two.  Further characteristic-three families with
boomerang uniformity zero or one were obtained in
\cite{KooKwonKoKimLocally2026,KooKwonKoKimAPN2026}.
Hu et al.~\cite{HuEtAl2023} determined the differential and boomerang
spectra of locally-APN power functions with boomerang uniformity two
in odd characteristic, while
Mesnager and Wu~\cite{MesnagerWu2025} studied locally-APN binomials
with boomerang uniformity at most two.
The locally-PN and locally-APN conditions used in these studies allow
exceptions to the corresponding differential bounds at certain output
differences; they do not imply that the
functions are PN or APN on the whole field.

For APN functions themselves, Pal and St\u{a}nic\u{a}~\cite{PalStanica2025}
related boomerang properties of odd APN functions to their
$(-1)$-differential properties.  Their results include APN inverse
functions with boomerang uniformity two, for example over $\F_{5^n}$
for every odd $n$.  For functions with differential uniformity at most
two over a field of odd characteristic, boomerang uniformity zero is
equivalent to perfect nonlinearity
\cite[Proposition~3.9 and Corollary~3.10]{KooKwonKoKim2026}.
Thus every APN function has $\beta_F\ge1$,
and constructing APN functions with $\beta_F=1$ attains the smallest
possible boomerang uniformity within the APN class.

Our previous work~\cite{KooKwonKoKim2026} gives such infinite families
from Dembowski--Ostrom (DO) PN functions over $\F_{3^n}$ with odd
$n>1$.  It uses sign-switching and affine perturbations and determines
the resulting differential and boomerang spectra.  Its CCZ-inequivalence
results follow from comparisons of differential spectra and from showing
that CCZ-equivalence of the
sign-switched functions forces EA-equivalence of the underlying PN functions.
The present paper develops a different, one-point modification
construction that works in every odd characteristic and applies to
every even PN function, including functions not represented by DO
polynomials.  It also gives exact formulas for the differential and
boomerang spectra and a criterion for CCZ-equivalence within the
construction, leading to many
pairwise CCZ-inequivalent classes with boomerang uniformity one.

Let $q=p^n>3$, let $g:\F_q\to\F_q$ be even and PN with $g(0)=0$,
and let $M$ be an $\F_p$-linearized polynomial.  For
$\lambda\in\F_q^*$ satisfying $-\lambda\notin g(\F_q)$, define
\[
 G_{\lambda,M}(x)=g(x)+M(x)+\lambda(x^{q-1}-1).
\]
This function agrees with the PN function $g+M$ outside zero and
takes the value $-\lambda$ at zero.  Thus it is a one-point
modification of a PN function EA-equivalent to $g$.

\paragraph{Main contributions.}
In Section~\ref{subsec:construction-spectra},
Theorem~\ref{thm:even-PN-onepoint} proves that every $G_{\lambda,M}$
is APN and completely classifies when its boomerang uniformity is
one or two.  For every linearized permutation $M$, each of the
$(q-1)/2$ values $\lambda\in\F_q^*$ satisfying
$-\lambda\notin g(\F_q)$ yields an APN function $G_{\lambda,M}$
with boomerang uniformity one.  The theorem also identifies an exceptional nonpermutation
case in characteristic three.
Corollary~\ref{cor:even-PN-global} gives the exact DDT and BCT row
distributions and complete formulas for both spectra.  The differential
spectrum depends only on $q$; when $\beta_{G_{\lambda,M}}=1$, the
boomerang spectrum is expressed in terms of $q$ and the explicitly
defined count $z_{\lambda,M}$.

Section~\ref{subsec:spectrum-inequivalence} establishes novelty
relative to known APN families.  For $q>7$,
Theorem~\ref{cor:listed-monomial-inequivalence} shows that the common
differential spectrum differs from that of every APN monomial in
Table~\ref{tab:apn-power-functions} and every APN binomial in the
original Ness--Helleseth family~\cite{NessHelleseth2007} or its extension
by Zeng et al.~\cite{ZengHuYangJiang2007}.  Since the differential
spectrum is CCZ-invariant, our functions are CCZ-inequivalent to all
these functions.  They are also CCZ-inequivalent to the APN families
constructed in
\cite{KooKwonKoKim2026}.  Indeed, over $\F_{3^n}$ with odd $n>1$,
those families have a total of $4(q-3)$ zero entries in the DDT rows indexed
by nonzero elements~\cite[Theorem~3.2]{KooKwonKoKim2026}, whereas
our functions have $2(q-3)$ by Corollary~\ref{cor:even-PN-global}.

Section~\ref{sec:explicit-spectra} evaluates the boomerang spectra
in closed form for explicit choices of $g$ and $M$.
Section~\ref{sec:coulter-matthews} treats Coulter--Matthews PN
functions and considers both permutation and nonpermutation linearized
perturbations.  For $\lambda=1$, $M(x)=x$, and Coulter--Matthews exponent
$(3^k+1)/2$, the number of boomerang-spectrum types obtained over
$\F_{3^n}$ with odd $n>1$ satisfying $\gcd(k,n)=1$ is unbounded
as the odd positive integer $k$ varies.
Section~\ref{sec:DO-examples} gives a systematic evaluation for
every DO PN function $g$ by taking $M=L_c$, where
$L_c(x)=D_cg(x)-g(c)$ is its linearized derivative for $c\in\F_q^*$.
Proposition~\ref{prop:derivative-z}
determines $z_{\lambda,L_c}$ and the number of nonzero parameters $c$
yielding each possible value.  These counts depend only on $q$.
Corollary~\ref{cor:translated-derivative-family} gives the resulting
boomerang spectrum.

The common differential spectrum cannot distinguish members of our
construction over the same field.  Section~\ref{subsec:one-point-rigidity}
characterizes CCZ-equivalence within this construction.
Theorem~\ref{thm:onepoint-ccz-rigidity} proves that functions arising
from even PN functions $g,h$ satisfying $g(0)=h(0)=0$ and parameters
$\lambda,\mu\in\F_q^*$ satisfying $-\lambda\notin g(\F_q)$ and
$-\mu\notin h(\F_q)$ are CCZ-equivalent if and only if there are invertible
$\F_p$-linear maps $M_1,M_2$ such that
$h\circ M_1=M_2\circ g$ and $M_2(\lambda)=\mu$.
The criterion is independent of the linearized perturbations.
Consequently, Corollary~\ref{cor:inequivalent-cores} transfers linear
inequivalence of the underlying even PN functions to CCZ-inequivalence
of the constructed APN functions.

Section~\ref{subsec:PN-inequivalent-families} applies this criterion
using nucleus orders of semifields associated with DO PN functions,
known inequivalence results, and distinct algebraic degrees of
Coulter--Matthews functions.
Theorem~\ref{thm:three-odd-characteristic-classes} gives three pairwise
CCZ-inequivalent APN functions with boomerang uniformity one over
$\F_{p^{15\ell}}$ for every odd prime $p$ and $\ell\ge1$, and six over
$\F_{p^{120r}}$ for every $r\ge1$.
Corollary~\ref{cor:ternary-classes} gives $\varphi(15\ell)/2+4$ such
classes over $\F_{3^{15\ell}}$ when $\ell>1$ is odd, where $\varphi$
is Euler's totient function.  In particular, the number of classes
obtained in characteristic three is unbounded.

Section~\ref{subsec:one-point-distance} develops further consequences.
Corollary~\ref{cor:ccz-classes-near-PN}(ii) proves that every function
at Hamming distance at most $m$ from some PN function has its entire
CCZ class equal to its EA class whenever $m(m+1)<q-q/p$.
The collection of all such functions is itself CCZ-invariant and,
for $m\ge1$ satisfying this bound, properly contains the PN functions.
Corollary~\ref{cor:onepoint-maximal-degree} shows that every nontrivial
one-point modification of a PN function has maximal algebraic degree
$n(p-1)$; this applies to all our APN functions.
Finally, Corollary~\ref{cor:parameter-orbits} describes CCZ-equivalence
as $\lambda$ varies through the orbits of a group determined by $g$.
APN pairs at Hamming distance one already occur implicitly in the
construction of Zha and Hu~\cite{ZhaHu2013}.
In our construction, for each fixed even PN function $g$ and linearized
polynomial $M$, distinct parameters $\lambda,\mu\in\F_q^*$ with
$-\lambda,-\mu\notin g(\F_q)$ give such pairs.  Choosing $M$ to be
a permutation gives boomerang uniformity one for every member.  The monomial case in
Corollary~\ref{cor:monomial-parameter-equivalence} gives EA-equivalent
pairs, whereas Proposition~\ref{prop:DY-distance-one} gives a
CCZ-inequivalent pair from a Ding--Yuan function over $\F_{3^5}$.

The paper is organized as follows.  Section~\ref{sec:preliminaries}
collects the required definitions and preliminary results.
Sections~\ref{sec:onepoint-construction} and~\ref{sec:explicit-spectra}
develop the general construction and its explicit spectrum evaluations,
respectively.  Section~\ref{sec:ccz-rigidity} characterizes
CCZ-equivalence within the construction, derives CCZ-inequivalence
results, and develops further applications.
Section~\ref{sec:conclusion} concludes the paper and discusses further directions.

\Needspace{5\baselineskip}
\section{Preliminaries}\label{sec:preliminaries}

For a nonzero function $F:\F_q\to\F_q$, write
$F(x)=\sum_{i=0}^{q-1}c_i x^i$, where $c_i\in\F_q$.
For $i=\sum_{j=0}^{n-1}i_jp^j$, with $0\le i_j<p$, define its
$p$-weight by $w_p(i):=\sum_{j=0}^{n-1}i_j$.
The \emph{algebraic degree} of $F$, denoted by
$\deg_{\mathrm{alg}}F$, is defined by
\[
 \deg_{\mathrm{alg}}F:=\max_{c_i\ne0}w_p(i).
\]
We set $\deg_{\mathrm{alg}}0=0$.
Since $w_p(i)\le n(p-1)$ for $0\le i\le q-1$, one has
$\deg_{\mathrm{alg}}F\le n(p-1)$.
The unique exponent $0\le i\le q-1$ with $w_p(i)=n(p-1)$ is
$i=q-1$; hence $\deg_{\mathrm{alg}}F=n(p-1)$ holds precisely when $c_{q-1}\ne0$.

For a function $F:\F_q\to\F_q$, we call $F$ \emph{even} or
\emph{odd} if, respectively,
\[
 F(-x)=F(x)\quad\text{or}\quad F(-x)=-F(x)\qquad(x\in\F_q).
\]
Since $q$ is odd, every odd function satisfies $F(0)=0$.
Every linearized polynomial over $\F_q$ is odd.
A Dembowski--Ostrom (DO) polynomial over $\F_q$ has the
form~\cite{DembowskiOstrom1968}
\[
 g(x)=\sum_{0\le i\le j<n}a_{ij}x^{p^i+p^j},
 \qquad a_{ij}\in\F_q.
\]
Every DO polynomial is even and vanishes at zero.

Let $F:\F_q\to\F_q$ be any function.  For $a\in\F_q$, put
\[
 D_aF(x):=F(x+a)-F(x)
\]
and put
\[
 L_{a,F}(x)
 :=D_aF(x)-D_aF(0)
 =F(x+a)-F(x)-F(a)+F(0).
\]
When the underlying function is clear, we simply write $L_a$.  For a
general function $F$, the map $L_{a,F}$ need not be linear.  If $F$ is
quadratic, that is, $\deg_{\mathrm{alg}}F=2$, then $D_aF$ is affine and $L_{a,F}$ is its $\F_p$-linear
part; in this case, $L_{a,F}$ is called the \emph{linearized
derivative} of $F$ with increment $a$.

For $a,b\in\F_q$, define the entry of the difference distribution
table (DDT) of $F$ at $(a,b)$ by
\[
 \delta_F(a,b):=\DDT_F(a,b):=
 \#\{x\in\F_q:D_aF(x)=b\}.
\]
Following Nyberg~\cite{Nyberg94}, the differential uniformity of $F$ is
\[
 \delta_F:=\max_{a\in\F_q^*,\ b\in\F_q}\delta_F(a,b).
\]
The function $F$ is perfect nonlinear (PN), also called planar in odd
characteristic, when $\delta_F=1$, and it is almost perfect nonlinear
(APN) when $\delta_F=2$.
By a \emph{DO PN function}, we mean a PN function represented by a DO
polynomial.

The boomerang connectivity table was introduced for permutations by Cid
et al.~\cite{CidEtAl18} and was extended to arbitrary functions by Li
et al.~\cite{LiQuSunLi2019}.  For $a,b\in\F_q^*$, its entries and the
boomerang uniformity are
\[
 \beta_F(a,b):=\BCT_F(a,b):=
 \#\{(x,y)\in\F_q^2:
       F(y+a)-F(x+a)=b,\ F(y)-F(x)=b\}
\]
and
\[
 \beta_F:=\max_{a,b\in\F_q^*}\beta_F(a,b),
\]
respectively.

For fixed $a\in\F_q^*$ and $b\in\F_q$, the \emph{fiber of $D_aF$
over $b$} is
\[
 (D_aF)^{-1}(b)=\{x\in\F_q:D_aF(x)=b\}.
\]
Unless another map is explicitly specified, ``fiber'' refers to a
fiber of a derivative.  Thus $\delta_F(a,b)$ is
the cardinality of the fiber of $D_aF$ over $b$.
Subtracting the two equations defining $\beta_F(a,b)$ gives the equivalent
and particularly useful form
\begin{equation}\label{eq:bct-fiber-form}
 \beta_F(a,b)
 =\#\bigl\{(x,y)\in\F_q^2:
       D_aF(x)=D_aF(y),\ F(y)-F(x)=b\bigr\}.
\end{equation}

For fixed $a\in\F_q^*$, the $a$-row of the DDT consists of the entries
$\delta_F(a,b)$ indexed by $b\in\F_q$, whereas the $a$-row of the BCT
consists of the entries $\beta_F(a,b)$ indexed by $b\in\F_q^*$.  In
either table, $b$ is the column index.  We use exponent notation for
the value distribution of a row: it has type
$0^{m_0}1^{m_1}\cdots$ if it contains exactly $m_i$ entries equal to
$i$; terms with $m_i=0$ are omitted.

For a function $F:\F_q\to\F_q$ and $i\ge0$, define the DDT and BCT
entry counts by
\begin{align*}
 N_i^{\DDT}(F)
   &=\#\{(a,b)\in\F_q^*\times\F_q:\delta_F(a,b)=i\},\\
 N_i^{\BCT}(F)
   &=\#\{(a,b)\in\F_q^*\times\F_q^*:\beta_F(a,b)=i\}.
\end{align*}
Using $v^m$ for $m$ copies of $v$, with $v^0$ omitted, define the
differential and boomerang spectra of $F$ by the multisets
\[
\begin{aligned}
 \mathcal S_{\DDT}(F)
   &:=\{\!\{0^{N_0^{\DDT}(F)},\ldots,
        \delta_F^{N_{\delta_F}^{\DDT}(F)}\}\!\},\\
 \mathcal S_{\BCT}(F)
   &:=\{\!\{0^{N_0^{\BCT}(F)},\ldots,
        \beta_F^{N_{\beta_F}^{\BCT}(F)}\}\!\}.
\end{aligned}
\]

We recall the following property of even PN functions; see also
K\"olsch and Polujan~\cite[Proposition~6.3]{KolschPolujan2024}.
We include a short proof.

\begin{lemma}\label{lem:even-PN-preimages}
Let $q$ be an odd prime power, and let $g:\F_q\to\F_q$ be an even PN
function satisfying $g(0)=0$.  Then $g$ is two-to-one:
$g^{-1}(0)=\{0\}$ and every nonzero value in its image has exactly
two preimages.  In particular, $|g(\F_q)|=(q+1)/2$.
\end{lemma}

\begin{proof}
Evenness gives $g(x)=g(-x)$.  Suppose that
$g(x)=g(y)$ with $x\ne y$, and put $a=x-y$.  Then
\[
                         D_ag(y)=0=D_ag(-a/2),
\]
where the second equality follows from evenness.  Since $g$ is PN,
$D_ag$ is a permutation, so $y=-a/2$ and hence $x=-y$.
Thus $g(x)=g(y)$ if and only if $y=\pm x$, and the assertions follow.
\end{proof}

For DO polynomials, Coulter and Matthews~\cite[Theorem~3]{CoulterMatthews2011}
proved the converse as well: a DO polynomial is PN if and only if it is
two-to-one, equivalently, if and only if its image has size $(q+1)/2$.

Throughout the paper, $\chi$ denotes the quadratic character of $\F_q$,
extended by $\chi(0)=0$.  For an even PN function $g:\F_q\to\F_q$
with $g(0)=0$, we write $g^{-1}(t):=\{x\in\F_q:g(x)=t\}$ for the
preimage set of $t\in\F_q$.  By Lemma~\ref{lem:even-PN-preimages},
$|g^{-1}(0)|=1$ and $|g^{-1}(t)|\in\{0,2\}$ for $t\ne0$.

\begin{lemma}
\label{lem:character-preimage-counts}
Let $g:\F_q\to\F_q$ be an even PN function satisfying $g(0)=0$,
where $q=p^n>3$ is odd.

\smallskip
\noindent\textup{(i)} For every $t\in\F_q$,
\[
 |g^{-1}(-t)|=\chi(-1)|g^{-1}(t)|+1-\chi(-1).
\]

\noindent\textup{(ii)} For every $\lambda\in\F_q^*$ such that
$-\lambda\notin g(\F_q)$,
\[
 |g^{-1}(\lambda)|=1-\chi(-1),\qquad
 |g^{-1}(-\lambda/2)|=1-\chi(2).
\]
\end{lemma}

\begin{proof}
\textup{(i)} For $t\ne0$, planarity gives
\begin{equation}\label{eq:PN-difference-count}
 \#\{(x,y)\in\F_q^2:g(x)-g(y)=t\}=q-1.
\end{equation}
Indeed, for each $a\in\F_q^*$, the equation $D_ag(y)=t$ has a unique
solution $y$.  As $a$ varies, these solutions yield the $q-1$ ordered
pairs $(x,y)=(y+a,y)$ counted in \eqref{eq:PN-difference-count}.
Among these pairs, $|g^{-1}(t)|+|g^{-1}(-t)|$ have exactly one zero
coordinate.  All remaining pairs form orbits of size four under
independent sign changes, since $g$ is even.  Hence
$|g^{-1}(t)|+|g^{-1}(-t)|\equiv q-1\pmod4$.  Since
$q-1\equiv1-\chi(-1)\pmod4$ and
$|g^{-1}(t)|,|g^{-1}(-t)|\in\{0,2\}$, we obtain
\textup{(i)} for $t\ne0$; the case $t=0$ follows from
$|g^{-1}(0)|=1$.

\smallskip
\noindent\textup{(ii)} Now let $\lambda\in\F_q^*$ satisfy
$-\lambda\notin g(\F_q)$.  Taking $t=-\lambda$ in \textup{(i)}
and using $|g^{-1}(-\lambda)|=0$ gives
$|g^{-1}(\lambda)|=1-\chi(-1)$.

For $t\ne0$, we have
\[
\begin{aligned}
 \#\{(x,y)\in\F_q^2:g(x)+g(y)=t\}
 &=\sum_{x\in\F_q}\bigl|g^{-1}(t-g(x))\bigr|\\
 &=\chi(-1)\sum_{x\in\F_q}\bigl|g^{-1}(g(x)-t)\bigr|
   +(1-\chi(-1))q\\
 &=\chi(-1)(q-1)+(1-\chi(-1))q=q-\chi(-1).
\end{aligned}
\]
Here the second equality follows from \textup{(i)}, applied to
$g(x)-t$, and the third equality follows from
\eqref{eq:PN-difference-count}.

Put $m=|g^{-1}(-\lambda/2)|$.  No solution $(x,y)\in\F_q^2$ of
$g(x)+g(y)=-\lambda$ has a zero coordinate, since
$-\lambda\notin g(\F_q)$.  There are exactly $2m$ such solutions
satisfying $x=\pm y$.  Since $g$ is even, each remaining solution
$(x,y)$, with $x\ne\pm y$, belongs to an orbit consisting of the
eight distinct solutions $(\pm x,\pm y)$ and $(\pm y,\pm x)$,
where the signs are chosen independently.  Consequently,
\[
 2m\equiv q-\chi(-1)
 \equiv2\bigl(1-\chi(2)\bigr)\pmod8,
\]
where the second congruence follows from the character values according
to $q\bmod8$.  By Lemma~\ref{lem:even-PN-preimages}, $m\in\{0,2\}$, so the congruence yields $m=1-\chi(2)$.
\end{proof}

\subsection*{Equivalence relations and graph intersections}

For a function $F:\F_q\to\F_q$, write
\[
             \mathcal G_F:=\{(x,F(x)):x\in\F_q\}
             \subseteq\F_q\times\F_q
\]
for its graph.  For $v=(a,b)\in\F_q\times\F_q$, write
$\delta_F(v):=\delta_F(a,b)$.  Then
\begin{equation}\label{eq:graph-intersection}
 \delta_F(v)
 =|\mathcal G_F\cap(\mathcal G_F-v)|
 =|\mathcal G_F\cap(\mathcal G_F+v)|.
\end{equation}

Two functions $F,F':\F_q\to\F_q$ are
\emph{Carlet--Charpin--Zinoviev (CCZ) equivalent}~\cite{CCZ98}, written
$F\sim_{\CCZ}F'$, if an affine automorphism of the $\F_p$-vector space
$\F_q\times\F_q$ maps $\mathcal G_F$ onto $\mathcal G_{F'}$.  They are
\emph{EA-equivalent}, written $F\sim_{\EA}F'$, if there exist invertible
$\F_p$-linear maps $M_1,M_2:\F_q\to\F_q$, an $\F_p$-linear map
$B:\F_q\to\F_q$, and constants $a_0,b_0\in\F_q$ such that
\[
 F'(M_1(x)+a_0)=M_2(F(x))+B(x)+b_0
 \qquad(x\in\F_q).
\]
EA-equivalence is a special case of CCZ-equivalence, since the affine
automorphism $(x,y)\mapsto(M_1(x)+a_0,\,M_2(y)+B(x)+b_0)$ maps
$\mathcal G_F$ onto $\mathcal G_{F'}$.

They are \emph{linearly equivalent} if $F'(M_1(x))=M_2(F(x))$ for
invertible $\F_p$-linear maps $M_1,M_2$.

\begin{lemma}\label{lem:ddt-entry-transport}
Suppose that the affine automorphism
$\mathcal A(P)=\mathcal M(P)+\mathbf c$, where $\mathcal M$ is the
$\F_p$-linear part of $\mathcal A$, maps $\mathcal G_F$ onto
$\mathcal G_{F'}$.  Then
\[
                    \delta_F(v)=\delta_{F'}(\mathcal M(v))
                    \qquad(v\in\F_q\times\F_q).
\]
\end{lemma}

\begin{proof}
By \eqref{eq:graph-intersection}, $\mathcal A$ restricts to a bijection
\[
 \mathcal G_F\cap(\mathcal G_F-v)
 \longrightarrow
 \mathcal G_{F'}\cap(\mathcal G_{F'}-\mathcal M(v)).
\]
Taking cardinalities proves the assertion.
\end{proof}

\section[A new construction of APN functions with boomerang uniformity one in odd characteristic]{A new construction of APN functions with\\boomerang uniformity one in odd characteristic}
\label{sec:onepoint-construction}

By~\cite[Corollary~3.10]{KooKwonKoKim2026}, every APN function
$F:\F_q\to\F_q$ over a finite field of odd characteristic satisfies
$\beta_F\ge1$.  The construction developed in this section attains
this minimum using any even PN function over $\F_q$, where
$q=p^n>3$ and $p$ is any odd prime.  The PN function need not be
represented by a DO polynomial; this extends the approach
in~\cite{KooKwonKoKim2026} for DO PN functions over $\F_{3^n}$
with odd $n>1$.
To the best of our knowledge, this is the first general construction
of APN functions with boomerang uniformity one valid in arbitrary
odd characteristic.

\subsection[Construction of new APN functions and their differential and boomerang spectra]{Construction of new APN functions and their\\differential and boomerang spectra}
\label{subsec:construction-spectra}

For a condition $P$, write $\mathbf1_{\{P\}}$ for its indicator function;
thus, $\mathbf1_{\{P\}}=1$ if $P$ holds and $\mathbf1_{\{P\}}=0$ otherwise.

Let $p$ be an odd prime, let $n\ge1$, and put $q=p^n>3$.
Let $g:\F_q\to\F_q$ be an even PN
function satisfying $g(0)=0$.  The normalization $g(0)=0$ can always
be achieved by subtracting $g(0)$, which preserves evenness and planarity.
Choose a constant $\lambda$ and an $\F_p$-linearized
polynomial $M$ satisfying
\begin{equation}\label{eq:onepoint-parameters}
 \lambda\in\F_q^*,\qquad -\lambda\notin g(\F_q),\qquad
 M(x)=\sum_{i=0}^{n-1}m_i x^{p^i},\quad m_i\in\F_q.
\end{equation}
Define
\begin{equation}\label{eq:onepoint-family}
 G_{\lambda,M}(x):=g(x)+M(x)+\lambda(x^{q-1}-1)
 =
 \begin{cases}
  -\lambda,&x=0,\\
  g(x)+M(x),&x\ne0.
 \end{cases}
\end{equation}
Thus $G_{\lambda,M}$ is obtained from $g+M$ by changing only its value at zero.

Unless the underlying function is explicitly specified otherwise, write
\begin{equation}\label{eq:core-centered-derivative}
 L_a(x):=L_{a,g}(x)=g(x+a)-g(x)-g(a).
\end{equation}
Then $L_a(x)=L_x(a)$, $L_a(0)=0$, and $L_a(-a)=-2g(a)$.
Planarity makes $L_a$ a permutation whenever $a\ne0$, although
it need not be linear.

For the sets and counts introduced below, we omit the underlying PN
function from the subscripts when it is clear from the context, and
include it when necessary.  For $a\in\F_q^*$, put
\begin{equation}\label{eq:onepoint-direction-sets}
 \begin{aligned}
 s_a&:=L_a^{-1}(\lambda),\\
 \mathcal E_{\lambda}&:=g^{-1}(-\lambda/2)=\{a\in\F_q^*:2g(a)=-\lambda\},\\
 \mathcal D_{\lambda}&:=\F_q^*\setminus\mathcal E_{\lambda}.
 \end{aligned}
\end{equation}
The identity $L_a(-a)=-2g(a)$ shows that
$a\in\mathcal E_{\lambda}$ if and only if $s_a=-a$.
The set $\mathcal E_{\lambda}$ is exceptional in the sense that, as
shown in the proof of Theorem~\ref{thm:even-PN-onepoint} below, it
consists precisely of the nonzero $a\in\F_q$ for which
$D_aG_{\lambda,M}$ is a permutation.

\begin{theorem}[Differential and boomerang uniformity]
\label{thm:even-PN-onepoint}
Let $q=p^n>3$, where $p$ is an odd prime and $n\ge1$.
Let $g:\F_q\to\F_q$ be an even PN function satisfying $g(0)=0$.
Let $M$ be an $\F_p$-linearized polynomial over $\F_q$, and choose
$\lambda\in\F_q^*$ such that $-\lambda\notin g(\F_q)$.
Then the function $G=G_{\lambda,M}$ defined in
\eqref{eq:onepoint-family} satisfies $\delta_G=2$ and
\[
 \beta_G=
 \begin{cases}
  1,&M\text{ is a linearized permutation polynomial},\\[1mm]
  1,&p=3,\ n\text{ is odd},\
       \ker M=\{0\}\cup\mathcal E_{\lambda},\\[1mm]
  2,&\text{otherwise}.
 \end{cases}
\]
In particular, for each fixed linearized permutation polynomial $M$,
exactly $(q-1)/2$ values $\lambda\in\F_q^*$ satisfy
$-\lambda\notin g(\F_q)$, and every such value gives an APN function
$G_{\lambda,M}$ with boomerang uniformity one.
\end{theorem}

\begin{proof}
By Lemma~\ref{lem:even-PN-preimages}, since $-\lambda/2\ne0$, we have
$|\mathcal E_{\lambda}|=|g^{-1}(-\lambda/2)|\in\{0,2\}$.  Since $q>3$, it follows that
$|\mathcal D_{\lambda}|=q-1-|\mathcal E_{\lambda}|\ge q-3>0$.

Fix $a\in\F_q^*$.  The map $x\mapsto D_ag(x)+M(a)$ is a
permutation and agrees with $D_aG$ outside $\{0,-a\}$.  At these
two inputs, their values are
\[
\begin{array}{c|cc}
 x&0&-a\\ \hline
 D_ag(x)+M(a)&M(a)+g(a)&M(a)-g(a)\\
 D_aG(x)&M(a)+g(a)+\lambda&M(a)-g(a)-\lambda.
\end{array}
\]
If $a\in\mathcal E_{\lambda}$, then $\lambda=-2g(a)$, so the two values
are exchanged.  Hence $D_aG$ is a permutation, and \eqref{eq:bct-fiber-form} gives
\begin{equation}\label{eq:exceptional-table-entries}
 \delta_G(a,b)=1\quad(b\in\F_q),\qquad
 \beta_G(a,b)=0\quad(b\in\F_q^*).
\end{equation}

Suppose that $a\in\mathcal D_{\lambda}$.  Since $g(a)\ne0$ and
$g(a)+\lambda\ne0$, each of the sets
\[
 \{M(a)\pm g(a)\},\qquad \{M(a)\pm(g(a)+\lambda)\}
\]
has two elements.  They are disjoint because an intersection would
force $\lambda=0$ or $\lambda=-2g(a)$.  The derivative comparison therefore gives
\begin{equation}\label{eq:omitted-DDT-values}
 \delta_G(a,b)=
 \begin{cases}
  0,&b\in\{M(a)\pm g(a)\},\\
  2,&b\in\{M(a)\pm(g(a)+\lambda)\},\\
  1,&\text{otherwise}.
 \end{cases}
\end{equation}
Thus the two omitted values of $D_aG$ are exactly $M(a)\pm g(a)$.  Since
$\mathcal D_{\lambda}$ is nonempty, \eqref{eq:omitted-DDT-values}
also gives $\delta_G=2$.

Continue to assume that $a\in\mathcal D_{\lambda}$.
By the definition of $s_a$ in \eqref{eq:onepoint-direction-sets}, we have
$L_a(s_a)=\lambda\ne0=L_a(0)$, and hence $s_a\ne0$.  Moreover,
$s_a\ne-a$, since otherwise $\lambda=L_a(-a)=-2g(a)$, which would imply
$a\in\mathcal E_{\lambda}$.  Hence $s_a\notin\{0,-a\}$.  Therefore
\[
 D_aG(0)=g(a)+M(a)+\lambda=D_ag(s_a)+M(a)=D_aG(s_a).
\]
Thus $\{0,s_a\}$ is the two-element fiber over $M(a)+g(a)+\lambda$.
Recall from Section~\ref{sec:preliminaries} that the fiber of $D_aG$
over $b\in\F_q$ is the preimage set
$(D_aG)^{-1}(b)=\{x\in\F_q:D_aG(x)=b\}$.
Since $G-M$ is even and $M$ is $\F_p$-linear, we have
$G(-x)=G(x)-2M(x)$.  Hence the involution $\iota_a(x)=-x-a$ satisfies
\[
 D_aG(\iota_a(x))=G(-x)-G(-x-a)=2M(a)-D_aG(x).
\]
It therefore maps the fiber $\{0,s_a\}$ bijectively onto the fiber
$\{-a,-a-s_a\}$ over $M(a)-g(a)-\lambda$.  By \eqref{eq:omitted-DDT-values},
these are precisely the two nonsingleton fibers:
\begin{equation}\label{eq:nonsingleton-fibers}
                   \{0,s_a\},\qquad\{-a,-a-s_a\}.
\end{equation}

For $a\in\mathcal D_{\lambda}$, orient each two-element fiber in
\eqref{eq:nonsingleton-fibers} from its first listed point to its
second.  The corresponding $G$-differences are
\begin{align*}
 b_+(a)&:=g(s_a)+\lambda+M(s_a)=G(s_a)-G(0),\\
 b_-(a)&:=g(s_a)+\lambda-M(s_a)=G(-a-s_a)-G(-a).
\end{align*}
For the second identity, we used
$g(a+s_a)-g(a)=g(s_a)+L_a(s_a)=g(s_a)+\lambda$.  We have
\[
 b_+(a)-b_-(a)=2M(s_a).
\]
Moreover,
\[
 b_+(a)+b_-(a)=2(g(s_a)+\lambda)\ne0,
\]
because $-\lambda\notin g(\F_q)$.  Thus $b_+(a)$ and $b_-(a)$ cannot
be negatives of one another, and at most one can vanish.
By \eqref{eq:bct-fiber-form}, the ordered pairs $(0,s_a)$ and $(s_a,0)$
contribute to the BCT entries in the columns indexed by $b_+(a)$ and
$-b_+(a)$, respectively, whenever these indices are nonzero.  Likewise,
$(-a,-a-s_a)$ and $(-a-s_a,-a)$ contribute to the BCT entries in the
columns indexed by $b_-(a)$ and $-b_-(a)$, respectively, whenever these
indices are nonzero.  Thus, for $a\in\mathcal D_{\lambda}$ and
$b\in\F_q^*$,
\[
 \beta_G(a,b)
 =\sum_{\varepsilon\in\{1,-1\}}
   \left(\mathbf1_{\{b=\varepsilon b_+(a)\}}
        +\mathbf1_{\{b=\varepsilon b_-(a)\}}\right),
\]
where repeated BCT column indices are counted with multiplicity.

Equivalently, for $a\in\mathcal D_{\lambda}$ and $b\in\F_q^*$,
\begin{equation}\label{eq:onepoint-bct-piecewise}
 \beta_G(a,b)=
 \begin{cases}
  2,&M(s_a)=0\ \text{and}\
       b\in\{\pm(g(s_a)+\lambda)\},\\[1mm]
  1,&M(s_a)\ne0\ \text{and}\
       b\in\{\pm b_+(a),\pm b_-(a)\}\setminus\{0\},\\[1mm]
  0,&\text{otherwise}.
 \end{cases}
\end{equation}

Finally, for every $a\in\F_q^*$, we have $s_a\ne0$ and
\[
 L_{s_a}(a)=L_a(s_a)=\lambda.
\]
Since $L_{s_a}$ is a permutation, the definition of $s_{s_a}$ gives
\begin{equation}\label{eq:direction-involution}
                         s_{s_a}=a\qquad(a\in\F_q^*).
\end{equation}
Thus $a\mapsto s_a$ is an involution of $\F_q^*$.  On $\mathcal E_{\lambda}$ it sends $a$ to
$-a$, and hence it preserves both $\mathcal E_{\lambda}$ and
$\mathcal D_{\lambda}$.  Consequently, a BCT row containing an entry
equal to $2$ exists precisely when $\ker M\cap\mathcal D_{\lambda}$
is nonempty.  Since $\mathcal D_{\lambda}\ne\varnothing$ and every
BCT row indexed by this set contains at least one positive entry,
$\beta_G\in\{1,2\}$ and
\[
 \beta_G=2
 \quad\Longleftrightarrow\quad
 \ker M\cap\mathcal D_{\lambda}\ne\varnothing.
\]

If $M$ is a permutation, this intersection is empty, so $\beta_G=1$.
Suppose that $M$ is not a permutation and $\beta_G=1$.
Then $\ker M\subseteq\{0\}\cup\mathcal E_{\lambda}$.
Since $\ker M$ is a nonzero $\F_p$-subspace and
$|\mathcal E_{\lambda}|\le2$, this forces $p=3$, $|\ker M|=3$,
and $\ker M=\{0\}\cup\mathcal E_{\lambda}$.
The character evaluation in Lemma~\ref{lem:character-preimage-counts}(ii)
gives $|\mathcal E_{\lambda}|=|g^{-1}(-\lambda/2)|=1-\chi(2)$.
In characteristic $3$, $\chi(2)=(-1)^n$, so
$|\mathcal E_{\lambda}|=2$ implies that $n$ is odd.
Conversely, the stated exceptional kernel has empty intersection with
$\mathcal D_{\lambda}$ and therefore gives $\beta_G=1$.
This proves the classification.

Finally, Lemma~\ref{lem:even-PN-preimages} gives
$|g(\F_q)|=(q+1)/2$.  Since $0\in g(\F_q)$, exactly $(q-1)/2$
values $\lambda\in\F_q^*$ satisfy $-\lambda\notin g(\F_q)$.
\end{proof}

To describe the DDT and BCT row distributions and the corresponding
differential and boomerang spectra, define
\begin{equation}\label{eq:onepoint-row-counts}
 e_{\lambda}:=|\mathcal E_{\lambda}|,\qquad
 d_{\lambda}:=|\mathcal D_{\lambda}|=q-1-e_{\lambda},
\end{equation}
and
\begin{equation}\label{eq:onepoint-bct-counts}
 \begin{aligned}
 r_{\lambda,M}&:=|\ker M\cap\mathcal D_{\lambda}|,\\
 z_{\lambda,M}
 &:=\#\{s\in\mathcal D_{\lambda}:
          M(s)\in\{\pm(g(s)+\lambda)\}\}.
 \end{aligned}
\end{equation}

\begin{lemma}
\label{lem:z-root-count}
Let $q=p^n>3$, where $p$ is an odd prime and $n\ge1$.
Let $g:\F_q\to\F_q$ be an even PN function satisfying $g(0)=0$,
and let $M:\F_q\to\F_q$ be an $\F_p$-linear function.
Suppose that $\lambda\in\F_q^*$ satisfies $-\lambda\notin g(\F_q)$.
With the notation in \eqref{eq:onepoint-direction-sets} and
\eqref{eq:onepoint-row-counts}--\eqref{eq:onepoint-bct-counts}, we have
\begin{equation}\label{eq:onepoint-character-counts}
 e_{\lambda}=1-\chi(2),\qquad d_{\lambda}=q-2+\chi(2).
\end{equation}
Moreover,
\[
 z_{\lambda,M}
 =2\,\#\{s\in\mathcal D_{\lambda}:g(s)+M(s)+\lambda=0\}.
\]
In particular, $z_{\lambda,M}$ is even.
\end{lemma}

\begin{proof}
Since $-\lambda/2\ne0$ and $g(0)=0$, Lemma~\ref{lem:character-preimage-counts}(ii)
gives $e_{\lambda}=|g^{-1}(-\lambda/2)|=1-\chi(2)$.
Together with $d_{\lambda}=q-1-e_{\lambda}$, this proves
\eqref{eq:onepoint-character-counts}.

Evenness of $g$ and oddness of $M$ give
$g(-s)+M(-s)+\lambda=g(s)-M(s)+\lambda$ for every $s\in\F_q$.
Since $\mathcal D_{\lambda}$ is closed under negation, $s\mapsto-s$
is a bijection between the sets
\[
 S_\pm:=\{s\in\mathcal D_{\lambda}:g(s)\pm M(s)+\lambda=0\}.
\]
These sets are disjoint: a common element would satisfy
$2(g(s)+\lambda)=0$, contradicting $-\lambda\notin g(\F_q)$.
By \eqref{eq:onepoint-bct-counts},
$z_{\lambda,M}=|S_+|+|S_-|=2|S_+|$.
\end{proof}

\Needspace{12\baselineskip}
\begin{corollary}[Row distributions and spectra]\label{cor:even-PN-global}
Under the hypotheses of Theorem~\ref{thm:even-PN-onepoint}, use the
notation \eqref{eq:onepoint-row-counts}--\eqref{eq:onepoint-bct-counts}.

\smallskip
\noindent\textup{(i) Differential distribution.}
The DDT row distribution is
\[
\begin{array}{c|c}
 \text{number of rows}&\text{DDT row type}\\ \hline
 e_{\lambda}&1^q\\
 d_{\lambda}&0^2 1^{q-4}2^2.
\end{array}
\]
Its differential spectrum is
\[
 \mathcal S_{\DDT}(G)
 =\{\!\{0^{2q-4+2\chi(2)},
          1^{q^2-5q+8-4\chi(2)},
          2^{2q-4+2\chi(2)}\}\!\}.
\]
Thus the differential spectrum depends only on $q$.

\smallskip
\noindent\textup{(ii) Boomerang distribution.}
The integers $r_{\lambda,M}$ and $z_{\lambda,M}$ are even.
The BCT row distribution is
\[
\begin{array}{c|c}
 \text{number of rows}&\text{BCT row type}\\ \hline
 e_{\lambda}&0^{q-1}\\
 r_{\lambda,M}&0^{q-3}2^2\\
 z_{\lambda,M}&0^{q-3}1^2\\
 d_{\lambda}-r_{\lambda,M}-z_{\lambda,M}&0^{q-5}1^4.
\end{array}
\]
Its boomerang spectrum is
\[
 \mathcal S_{\BCT}(G)=\{\!\{0^{(q-1)^2-4d_{\lambda}+2r_{\lambda,M}+2z_{\lambda,M}},
  1^{4d_{\lambda}-4r_{\lambda,M}-2z_{\lambda,M}},2^{2r_{\lambda,M}}\}\!\}.
\]
In particular, $\beta_G=1$ if and only if $r_{\lambda,M}=0$; in
this case, the boomerang spectrum is
\[
 \mathcal S_{\BCT}(G)
 =\{\!\{0^{q^2-6q+9-4\chi(2)+2z_{\lambda,M}},
          1^{4q-8+4\chi(2)-2z_{\lambda,M}}\}\!\}.
\]
\end{corollary}

\begin{proof}
By Lemma~\ref{lem:z-root-count}, we have $d_\lambda=q-2+\chi(2)$.
The DDT assertions follow from
\eqref{eq:exceptional-table-entries}--\eqref{eq:omitted-DDT-values}.
For the BCT, \eqref{eq:exceptional-table-entries} shows that the
$e_{\lambda}$ rows indexed by $a\in\mathcal E_{\lambda}$ are
identically zero.
The involution \eqref{eq:direction-involution} permutes
$\mathcal D_{\lambda}$.
By \eqref{eq:onepoint-bct-piecewise}, exactly $r_{\lambda,M}$ rows
have two entries equal to $2$, and exactly $z_{\lambda,M}$ rows
have two entries equal to $1$, the latter occurring when precisely
one of $b_+(a),b_-(a)$ vanishes.
Each of the remaining
$d_{\lambda}-r_{\lambda,M}-z_{\lambda,M}$ BCT rows
indexed by $\mathcal D_{\lambda}$ has four entries equal to $1$.
Summing these row types and substituting for $d_\lambda$ when
$r_{\lambda,M}=0$ gives the stated boomerang spectra.

The set $\ker M\cap\mathcal D_{\lambda}$ is closed under negation
and does not contain zero, so $r_{\lambda,M}$ is even.
The evenness of $z_{\lambda,M}$ follows from
Lemma~\ref{lem:z-root-count}.
\end{proof}

The common differential spectrum does not imply CCZ-equivalence among the
constructed functions.  Indeed, Theorem~\ref{thm:three-odd-characteristic-classes}
and Corollary~\ref{cor:ternary-classes} will provide explicit sets of pairwise
CCZ-inequivalent APN functions with boomerang uniformity one over the same
field, all sharing the same differential spectrum.

When $M$ is a permutation, $r_{\lambda,M}=0$ and
$\beta_G=1$.  At the other extreme, $M=0$ gives
$r_{\lambda,0}=d_{\lambda}$, $z_{\lambda,0}=0$, and
$\beta_G=2$.

\Needspace{7\baselineskip}
The condition $-\lambda\notin g(\F_q)$ in
\eqref{eq:onepoint-parameters} is sharp for the APN conclusion.  Indeed, if
$-\lambda\in g(\F_q)$, choose $a\ne0$ with $g(a)=-\lambda$.  Then
$s_a=-a/2$, and $D_aG_{\lambda,M}$ takes the same value at the three
distinct elements $0,-a/2,-a$.
Hence $G_{\lambda,M}$ is not APN.

If $\chi(-1)=-1$, Lemma~\ref{lem:character-preimage-counts}(i) gives
$|g^{-1}(-\lambda)|=2-|g^{-1}(\lambda)|$.  Thus, for every
$\lambda\in\F_q^*$,
\begin{equation}\label{eq:lambda-image-equivalence}
 -\lambda\notin g(\F_q)
 \quad\Longleftrightarrow\quad
 \lambda\in g(\F_q).
\end{equation}
Consequently, all values of $\lambda$ satisfying the condition in
\eqref{eq:onepoint-parameters} can be written as $\lambda=g(s)$,
with $s\in\F_q^*$.  By Lemma~\ref{lem:even-PN-preimages}, each value
arises from exactly the two elements $s$ and $-s$.

\paragraph{Nonpermutation linear perturbations.}

The nonpermutation case of Theorem~\ref{thm:even-PN-onepoint}
admits the following explicit realization.

\begin{corollary}
\label{cor:nonpermutation-linearized}
Let $q=3^n$, where $n>1$ is odd, and let $g:\F_q\to\F_q$ be an even
PN function satisfying $g(0)=0$.  Then, for every $c,u\in\F_q^*$, the function
\[
 G(x)=g(x)+u(x^3-c^2x)+g(c)(x^{q-1}-1)
\]
is APN with boomerang uniformity one.
\end{corollary}

\begin{proof}
Put $\lambda=g(c)$ and $M(x)=u(x^3-c^2x)$, so that $G=G_{\lambda,M}$.
Lemma~\ref{lem:even-PN-preimages} gives $\lambda=g(c)\ne0$.
Since $\chi(-1)=-1$, equivalence~\eqref{eq:lambda-image-equivalence}
gives $-\lambda\notin g(\F_q)$.
In characteristic $3$, the equation $2g(a)=-\lambda$ is equivalent
to $g(a)=g(c)$, so
$\mathcal E_{\lambda}=\{\pm c\}$ by
Lemma~\ref{lem:even-PN-preimages}.
The polynomial $M$ is $\F_3$-linearized, and the factorization
$M(x)=ux(x-c)(x+c)$ shows that
$\ker M=\{0,\pm c\}=\{0\}\cup\mathcal E_{\lambda}$.
Theorem~\ref{thm:even-PN-onepoint} now gives both uniformities.
\end{proof}

The function $g$ in Corollary~\ref{cor:nonpermutation-linearized} can be
chosen from any of the families $f_1,\ldots,f_5$ listed in
Table~\ref{tab:general-PN-families} of Section~\ref{sec:DO-examples},
with $p=3$, odd $n>1$, and the parameters satisfying the conditions
in the corresponding row.

\subsection[CCZ-inequivalence to known APN monomials and Ness--Helleseth-type binomials]{CCZ-inequivalence to known APN monomials\\and Ness--Helleseth-type binomials}
\label{subsec:spectrum-inequivalence}

We compare the differential spectrum in Corollary~\ref{cor:even-PN-global}(i)
with those of power functions and Ness--Helleseth-type binomials,
extending the divisibility argument in~\cite[Section~3.3]{KooKwonKoKim2026}.
For a positive integer $r$ and $u\in\F_q^*$, write
\[
 H_{r,u}(x):=x^r(1+u\chi(x)),
\]
and call $H_{r,u}$ a \emph{Ness--Helleseth-type binomial}.
The original Ness--Helleseth family is the special case
\[
 H_{q-2,u}(x)=x^{q-2}+ux^{(q-3)/2}
\]
as polynomial functions on $\F_q$, where $q=3^n$ and $n\ge3$ is odd.
Ness and Helleseth~\cite{NessHelleseth2007} proved that this function
is APN when $\chi(u-1)=\chi(u+1)=\chi(u)$.
Zeng et al.~\cite{ZengHuYangJiang2007} extended this construction to
$q=p^n$, where $p\ge7$, $p\equiv3\pmod4$, and $n$ is odd:
$H_{q-2,u}$ is APN if either
$\chi(u+1)=\chi(u-1)=-\chi(5u+3)$ or
$\chi(u+1)=\chi(u-1)=-\chi(5u-3)$.

\begin{proposition}\label{prop:spectrum-inequivalence}
Suppose that $q>5$ and $\chi(2)=-1$
(equivalently, $q\equiv\pm3\pmod8$).
Every function $G_{\lambda,M}$ satisfying the hypotheses of
Theorem~\ref{thm:even-PN-onepoint} has a differential spectrum different
from that of every power function and every Ness--Helleseth-type binomial
over $\F_q$.  Consequently, $G_{\lambda,M}$ is CCZ-inequivalent to
all such functions.
\end{proposition}

\begin{proof}
For a power function $H(x)=x^d$, where $d$ is a positive integer,
substitution $x=ay$ gives
\[
 \delta_H(a,b)=\delta_H(1,b/a^d)\qquad(a\ne0).
\]
Thus all DDT rows indexed by nonzero elements have the same multiset
of entries, and $q-1$ divides every $N_i^{\DDT}(H)$.
For $H=H_{r,u}$ and every nonzero square $c$, we have $H(cx)=c^rH(x)$,
whence
\[
 \delta_H(ca,c^rb)=\delta_H(a,b).
\]
For $a\in\F_q^*$, put
\begin{equation}\label{eq:DDT-row-multiplicities}
 \omega_i(a):=\#\{b\in\F_q:\delta_H(a,b)=i\}.
\end{equation}
Since multiplication by $c^r$ permutes $\F_q$, the scaling identity above
and the definition of $\omega_i(a)$ imply $\omega_i(ca)=\omega_i(a)$ for every nonzero square $c$.
Let $S$ be the set of nonzero squares and fix a nonsquare $k$.
Then $\F_q^*=S\sqcup kS$, with $|S|=|kS|=(q-1)/2$, and hence
\[
 N_i^{\DDT}(H)
 =\sum_{a\in\F_q^*}\omega_i(a)
 =\frac{q-1}{2}\bigl(\omega_i(1)+\omega_i(k)\bigr).
\]
Thus $(q-1)/2$ divides every $N_i^{\DDT}(H)$.
The rows indexed by $1$ and $k$ need not have the same distribution.

On the other hand, since $\chi(2)=-1$, equation
\eqref{eq:onepoint-character-counts} and
Corollary~\ref{cor:even-PN-global}(i) give
\[
 N_2^{\DDT}(G_{\lambda,M})
 =2\bigl(q-2+\chi(2)\bigr)=2(q-3)=4\frac{q-1}{2}-4.
\]
If $G_{\lambda,M}$ had the same differential spectrum as a power
function or a Ness--Helleseth-type binomial, then $q-1\mid8$,
which is impossible because $q>5$ and $q\equiv\pm3\pmod8$ imply
$q\ge11$.
Finally, Lemma~\ref{lem:ddt-entry-transport} implies that the
differential spectrum is CCZ-invariant.  Hence the functions
are CCZ-inequivalent.
\end{proof}

\begin{lemma}\label{lem:matching-power-spectrum}
Suppose that $\chi(2)=1$, and let $G_{\lambda,M}$ satisfy the
hypotheses of Theorem~\ref{thm:even-PN-onepoint}.
If a power function $F(x)=x^d$, with $d$ a positive integer,
has the same differential spectrum as $G_{\lambda,M}$,
then $d$ is even and
\[
                         q\le2d-1.
\]
\end{lemma}

\begin{proof}
Since $\chi(2)=1$, we have $e_\lambda=0$ and $d_\lambda=q-1$, so
Corollary~\ref{cor:even-PN-global}(i) shows that every nonzero DDT row
of $G_{\lambda,M}$ has type $0^2 1^{q-4}2^2$.
All nonzero DDT rows of a power function have the same distribution;
therefore, equality of the spectra forces every such row of $F$
to have this type.
If $d$ is odd, then $D_1F(-x-1)=D_1F(x)$.
The map $x\mapsto-x-1$ fixes only $-1/2$, and the remaining $q-1$
elements form $(q-1)/2$ pairs $\{x,-x-1\}$, on each of which $D_1F$
takes the same value.  The APN property implies that
the values of $D_1F$ on these pairs are all distinct and differ from
$D_1F(-1/2)$.  Thus the DDT row indexed by $1$ has type
$0^{(q-1)/2}1^1 2^{(q-1)/2}$, contradicting the required type
$0^2 1^{q-4}2^2$, since $q-4>1$ under the hypotheses $\chi(2)=1$ and $q>3$.
Thus $d$ is even.

Put $P(x)=(x+1)^d-x^d$.  The identity
$P(-x-1)=-P(x)$ gives $\delta_F(1,b)=\delta_F(1,-b)$.
Together with the row type $0^2 1^{q-4}2^2$, this implies that
there exist $u,v\in\F_q^*$ with $u^2\ne v^2$ such that
\[
 \delta_F(1,b)=
 \begin{cases}
  0,&b=\pm u,\\
  2,&b=\pm v,\\
  1,&b\notin\{\pm u,\pm v\}.
 \end{cases}
\]
Recall that $\sum_{x\in\F_q}1=q=0$ and
$\sum_{x\in\F_q}x^j=0$ for $1\le j<q-1$.
Consequently, every polynomial $R\in\F_q[x]$ of degree less than
$q-1$ satisfies $\sum_{x\in\F_q}R(x)=0$.
Since $q>3$, we obtain
\[
 \sum_{x\in\F_q}P(x)^2
 =\sum_{b\in\F_q}b^2-2u^2+2v^2
 =2(v^2-u^2)\ne0.
\]
Suppose, for a contradiction, that $q>2d-1$.
Then $\deg(P^2)\le2(d-1)<q-1$, so the power-sum identity gives
$\sum_{x\in\F_q}P(x)^2=0$, contradicting the preceding calculation.
Therefore $q\le2d-1$, proving the asserted bound.
\end{proof}

Table~\ref{tab:apn-power-functions} combines the odd-characteristic
families in~\cite[Table~1]{BudaghyanPal2025} with the ternary
families summarized in~\cite[Table~1]{KooKwonKoKim2026}.
The congruence description in row $F_{10}$ follows
\cite[Theorem~4.1]{ZhaWang2010} and
\cite[Section~3]{KooKwonKoKimAPN2026}.

\begingroup
\small
\setlength{\tabcolsep}{4pt}
\renewcommand{\arraystretch}{1.35}
\setlength{\LTcapwidth}{\textwidth}
\setlength{\LTpost}{6pt}
\begin{longtable}{@{}c
 >{\raggedright\arraybackslash}p{0.315\textwidth}
 c >{\raggedright\arraybackslash}p{0.395\textwidth} c@{}}
\caption{Known families of APN monomials $x^d$ over $\F_q$,
where $q=p^n>3$ and $p$ is odd.}
\label{tab:apn-power-functions}\\
\toprule
Family & Exponent $d$ & $p$ & Conditions & References \\
\midrule
\endfirsthead
\multicolumn{5}{c}{\small Table~\thetable\ (continued)}\\[3pt]
\toprule
Family & Exponent $d$ & $p$ & Conditions & References \\
\midrule
\endhead
\midrule
\multicolumn{5}{r}{\small Continued on the next page}\\
\endfoot
\bottomrule
\endlastfoot
$F_1$ & $3$ & $p\ne3$ & & \cite{HellesethRongSandberg1999} \\[4pt]
$F_2$ & $q-2$ & odd & $q\equiv2\pmod3$ & \cite{HellesethRongSandberg1999} \\[4pt]
$F_3$ & $\dfrac{q-3}{2}$ & odd
 & $p\equiv3,7\pmod{20}$, $n$ odd, $q>7$, $q\ne27$
 & \cite{HellesethRongSandberg1999} \\[4pt]
$F_4$ & $\dfrac{3q-1}{4}$ & odd
 & $q\equiv3\pmod8$ & \cite{HellesethRongSandberg1999} \\[4pt]
$F_5$ & $\dfrac{q+1}{4}$ & odd
 & $q\equiv7\pmod8$, $q>7$ & \cite{HellesethRongSandberg1999} \\[4pt]
$F_6$ & $p^m+2$ & odd
 & $m\ge1$, $n=2m$, $p^m\equiv1\pmod3$
 & \cite{HellesethRongSandberg1999} \\[4pt]
\midrule
$F_7$ & $q-3$ & $3$ & $n\ge3$ odd
 & \cite{HellesethRongSandberg1999} \\[4pt]
$F_8$ & $\dfrac{3^{(n+1)/2}-1}{2}+\epsilon_n\dfrac{q-1}{2}$ & $3$
 & $n\ge5$ odd & \cite{DobbertinEtAl2003,Leducq2012} \\[4pt]
$F_9$ & $\dfrac{3^{n+1}-1}{8}+\epsilon_n\dfrac{q-1}{2}$ & $3$
 & $n\ge5$ odd & \cite{DobbertinEtAl2003,Leducq2012} \\[4pt]
$F_{10}$ & $d(3^m+1)\equiv2\pmod{q-1}$,\newline $d$ even & $3$
 & $n\ge3$ odd, $1\le m\le(n-1)/2$, $\gcd(m,n)=1$
 & \cite{KooKwonKoKimAPN2026,ZhaWang2010} \\[4pt]
\midrule
$F_{11}$ & $\dfrac{5^k+1}{2}$ & $5$
 & $k\ge1$, $\gcd(2n,k)=1$ & \cite{HellesethRongSandberg1999} \\[4pt]
$F_{12}$ & $\dfrac{q-1}{4}+\dfrac{5^{(n+1)/2}-1}{2}$ & $5$
 & $n$ odd & \cite{DobbertinEtAl2003,ZhaWang2010} \\[4pt]
$F_{13}$ & $\dfrac{5^{n+1}-1}{2(5^{(n+1)/2^\ell}+1)}+\dfrac{q-1}{4}$ & $5$
 & $\ell\ge2$, $n\equiv-1\pmod{2^\ell}$
 & \cite{Leducq2012,ZhaWang2010} \\
\end{longtable}

\noindent
\begin{minipage}{\textwidth}
\small
In $F_8$ and $F_9$, $\epsilon_n=0$ for $n\equiv3\pmod4$ and
$\epsilon_n=1$ for $n\equiv1\pmod4$.

\smallskip
\noindent
In $F_{10}$, the congruence determines a unique even residue
modulo $q-1$.  The family $F_{10}$ includes the following previously
known exponents: $d=(3^{n+1}-1)/4$ by taking $m=1$, and
\[
 d=\frac{3^{n+1}-1}{3^{(n+1)/2^\ell}+1},
 \qquad \ell\ge2,\quad n\equiv-1\pmod{2^\ell},
\]
by taking $m=(n+1)/2^\ell$; see~\cite{KooKwonKoKimAPN2026,Leducq2012,ZhaWang2010}.
\end{minipage}
\par
\endgroup

\begin{theorem}[CCZ-inequivalence via differential spectra]\label{cor:listed-monomial-inequivalence}
Suppose that $q>7$.
Every function $G_{\lambda,M}$ satisfying the hypotheses of
Theorem~\ref{thm:even-PN-onepoint} has a differential spectrum
different from that of every APN monomial listed in
Table~\ref{tab:apn-power-functions} and every APN binomial over $\F_q$
in the original Ness--Helleseth family~\cite{NessHelleseth2007} or
its extension by Zeng et al.~\cite{ZengHuYangJiang2007}.
Consequently, it is CCZ-inequivalent to all these functions.
\end{theorem}

\begin{proof}
When $\chi(2)=-1$, the assertion follows from
Proposition~\ref{prop:spectrum-inequivalence}.
Suppose that $\chi(2)=1$, or equivalently $q\equiv\pm1\pmod8$.
We first consider the APN monomials in Table~\ref{tab:apn-power-functions}.
The families over $\F_{3^n}$ and $\F_{5^n}$
that require odd $n$, and the family requiring $q\equiv3\pmod8$,
cannot occur.
The remaining exponents in the table are
\[
 3,\quad q-2,\quad \frac{q-3}{2},\quad \frac{q+1}{4},
 \quad p^m+2,\quad \frac{5^k+1}{2}.
\]
The exponents $(q-3)/2$ and $(q+1)/4$ are strictly smaller than
$(q+1)/2$, whereas all the remaining exponents are odd.
Lemma~\ref{lem:matching-power-spectrum} therefore excludes
equality of the differential spectra for all the listed monomials.

We next consider the APN binomials.  The original ternary family cannot
occur when $\chi(2)=1$.  The extension by Zeng et al.\ has
$q\equiv3\pmod4$, so it remains to consider $q\equiv7\pmod8$.
Let $H=H_{q-2,u}$ be a member of that APN family, and use the notation
$\omega_i(a)$ from \eqref{eq:DDT-row-multiplicities}.
Since $q\equiv3\pmod4$, the result of Mesnager and
Wu~\cite{MesnagerWu2025} implies that all DDT rows of $H$ indexed by
nonzero elements have the same distribution.  Hence
\[
 N_2^{\DDT}(H)=(q-1)\omega_2(1),\qquad
 N_2^{\DDT}(G_{\lambda,M})=2(q-1).
\]
It suffices to show that $\omega_2(1)>2$.

Let $T(u)$ count the $b\in\F_q^*$ for which $H(x+1)-H(x)=b$
has exactly two solutions in $\F_q\setminus\{0,-1\}$.
By~\cite[Lemmas~4--5]{XiaLiBaoChenHelleseth2025}, this agrees with
the quantity $T(u)=T_1(u)+T_2(u)$ in Corollary~2
of~\cite{XiaLiBaoChenHelleseth2025}.
For each such $b$, the APN property of $H$ excludes any additional
solution at $0$ or $-1$, so $\delta_H(1,b)=2$.
Consequently, $\omega_2(1)\ge T(u)$.
The counting formula and Weil bound in~\cite[Corollary~2 and
Theorem~4]{XiaLiBaoChenHelleseth2025} yield
\[
 \omega_2(1)\ge T(u)\ge\frac{q-7}{4}-\sqrt q>2
 \qquad(q\ge47).
\]
The only prime powers with $7<q<47$ and $q\equiv7\pmod8$ are
$q=23$ and $q=31$.
For $q=23$, the parameters satisfying the conditions of Zeng et al.\
are $u\in\pm\{2,5,6,7\}$; for $q=31$, they are
$u\in\pm\{3,6,8,9,12,14\}$.
Direct evaluation of $H(x+1)-H(x)$ for these parameters gives
$\omega_2(1)\ge6$ in both fields.
Thus $N_2^{\DDT}(H)>N_2^{\DDT}(G_{\lambda,M})$ in every remaining
case, proving that the differential spectra differ.
CCZ-inequivalence follows from Lemma~\ref{lem:ddt-entry-transport}.
\end{proof}

\begin{remark}
\label{rem:spectrum-comparison-limits}
The restriction $q>5$ in Proposition~\ref{prop:spectrum-inequivalence}
cannot be omitted.  Over $\F_5$, the choice $g(x)=x^2$, $\lambda=2$, and
$M(x)=x$ gives $G_{2,M}(x)=H_{2,2}(x)+x-2$, where
$H_{2,2}(x)=x^2+2x^4$.  Therefore these two APN functions are EA-equivalent.

The restriction $q>7$ in Theorem~\ref{cor:listed-monomial-inequivalence}
is also essential for the differential-spectrum comparison.
Over $\F_7$, both the APN monomial $x^4$ and the APN binomial
$H_{5,3}(x)=x^5+3x^2$ from the extension by Zeng et al.~\cite{ZengHuYangJiang2007}
have nonzero DDT row type $0^2 1^3 2^2$.
Hence both have the same differential spectrum
$\{\!\{0^{12},1^{18},2^{12}\}\!\}$ as $G_{\lambda,M}$.

For power functions, Lemma~\ref{lem:matching-power-spectrum} shows that,
when $\chi(2)=1$, equality of the differential spectra of $x^d$ and
$G_{\lambda,M}$ requires $q\le2d-1$.
Consequently, for each fixed exponent $d$, equality is possible over
only finitely many fields.  In particular, for $d=4$, the only
possible field order is $q=7$.
The bound $q\le2d-1$ does not, however, exclude equality of the
differential spectra over infinitely many fields when $d$ is allowed
to depend on $q$.
\end{remark}

\section[Explicit APN functions with boomerang uniformity one and their spectra]{Explicit APN functions with boomerang uniformity one\\and their spectra}
\label{sec:explicit-spectra}

We apply the construction of Section~\ref{sec:onepoint-construction}
to Coulter--Matthews and DO PN functions to obtain explicit APN functions
with boomerang uniformity one.  For each fixed field $\F_q$, all APN functions
constructed in this section have the differential spectrum given in
Corollary~\ref{cor:even-PN-global}(i).  We therefore focus on their boomerang
spectra, obtaining closed formulas by evaluating $z_{\lambda,M}$ and,
in selected families, counting the parameter choices yielding each spectrum.

\subsection[APN functions with boomerang uniformity one from Coulter--Matthews PN functions]{APN functions with boomerang uniformity one\\from Coulter--Matthews PN functions}
\label{sec:coulter-matthews}

We apply the general construction of Section~\ref{sec:onepoint-construction}
to Coulter--Matthews PN functions.  This family includes functions not
represented by DO polynomials: for example, $x^{14}$ over $\F_{3^5}$
has algebraic degree $4$.  We first specialize the boomerang spectrum
formula of Corollary~\ref{cor:even-PN-global}(ii) to the Coulter--Matthews
family for both odd and even $n$.  For odd $n$, we then determine the
boomerang spectra for the nonpermutation choice
$M(x)=x^3-x$, for $M(x)=x$, and for a family of linearized permutation
binomials.

Coulter and Matthews~\cite[Theorem~4.1]{CoulterMatthews1997} proved that,
for $q=3^n$ with $n\ge2$ and $k\ge1$ satisfying $\gcd(k,2n)=1$, the
function
\[
                         g(x)=x^{(3^k+1)/2}
\]
is PN.  Since $k$ is odd, $g$ is even and $g(0)=0$.  Moreover,
$g(\F_q^*)$ is contained in the set of nonzero squares and has
$(q-1)/2$ elements by Lemma~\ref{lem:even-PN-preimages}; hence it is
exactly the set of nonzero squares.  For $\lambda\in\F_q^*$, we therefore have
\[
 -\lambda\notin g(\F_q)\quad\Longleftrightarrow\quad
 \chi(-\lambda)=-1\quad\Longleftrightarrow\quad
 \chi(\lambda)=(-1)^{n+1}.
\]
Thus $\lambda$ must be a nonzero square when $n$ is odd and a nonsquare
when $n$ is even.
For any $\lambda\in\F_q^*$ satisfying this condition and any linearized
polynomial $M$ over $\F_q$, define
\[
 G_{\lambda,M}(x)
 =x^{(3^k+1)/2}+M(x)+\lambda(x^{q-1}-1).
\]
Theorem~\ref{thm:even-PN-onepoint} therefore applies to $G_{\lambda,M}$.
Since $\chi(2)=(-1)^n$, equation~\eqref{eq:onepoint-character-counts} gives
\[
 e_{\lambda}=1-(-1)^n,\qquad
 d_{\lambda}=q-2+(-1)^n.
\]

For this family, write $z=z_{\lambda,M}$, with $z_{\lambda,M}$
defined in \eqref{eq:onepoint-bct-counts}.

\begin{corollary}
\label{cor:coulter-matthews}
Suppose that $M$ is a linearized permutation polynomial, or that
$n$ is odd and $\ker M=\{0\}\cup\mathcal E_{\lambda}$.
Then
$\delta_{G_{\lambda,M}}=2$, $\beta_{G_{\lambda,M}}=1$, and
\[
 \mathcal S_{\BCT}(G_{\lambda,M})
 =\{\!\{0^{q^2-6q+9-4(-1)^n+2z},
          1^{4q-8+4(-1)^n-2z}\}\!\}.
\]
\end{corollary}

\begin{proof}
In either case, Theorem~\ref{thm:even-PN-onepoint} gives
$\delta_{G_{\lambda,M}}=2$ and $\beta_{G_{\lambda,M}}=1$.
Corollary~\ref{cor:even-PN-global} therefore gives $r_{\lambda,M}=0$.
Substituting this and
$d_{\lambda}=q-2+(-1)^n$ into
Corollary~\ref{cor:even-PN-global}(ii) gives the boomerang spectrum.
\end{proof}

Corollary~\ref{cor:coulter-matthews} applies to both odd and even $n$.
Throughout the remainder of Section~\ref{sec:coulter-matthews}, we assume
that $n>1$ is odd.  We focus on obtaining explicit boomerang spectra by evaluating
$z_{\lambda,M}$ in closed form.

\subsubsection{The nonpermutation linear perturbation \texorpdfstring{$M(x)=x^3-x$}{M(x)=x³-x}}
\label{subsec:CM-nonpermutation}

By Theorem~\ref{thm:even-PN-onepoint}, a nonpermutation $M$ can give
boomerang uniformity one only when $n$ is odd.
For every odd $k\ge1$ with
$\gcd(k,n)=1$, the Coulter--Matthews function
$g(x)=x^{(3^k+1)/2}$ satisfies $g(1)=1$.
Corollary~\ref{cor:nonpermutation-linearized}, with $c=u=1$, therefore
gives an APN function with boomerang uniformity one by taking
$M(x)=x^3-x$ and $\lambda=1$.

To evaluate $z_{\lambda,M}$ and thereby obtain an explicit boomerang
spectrum with a nonpermutation $M$, take $k=n+2$.
The corresponding PN function is
$g(x)=x^{(q+9)/2}$ on $\F_q$, as verified below.

\begin{corollary}
\label{cor:nonpermutation-CM-spectra}
For $q=3^n$ with odd $n>1$, the function
\[
 G(x):=x^{(q+9)/2}+x^3-x+x^{q-1}-1
\]
satisfies $\delta_G=2$ and $\beta_G=1$.  Its boomerang spectrum is
\[
 \mathcal S_{\BCT}(G)
  =\{\!\{0^{q^2-6q+13},1^{4(q-3)}\}\!\}.
\]
\end{corollary}

\begin{proof}
Since $k=n+2$ is odd and $\gcd(k,n)=1$, the function
$x^{(3^k+1)/2}$ is PN by
Coulter and Matthews~\cite{CoulterMatthews1997}.
Moreover, $(3^k+1)/2=(9q+1)/2=(q+9)/2+4(q-1)$,
so this function is $g(x)=x^{(q+9)/2}$ on $\F_q$.
The function $g$ is even and vanishes at zero, and our choices satisfy
$M(x)=x^3-x$ and $\lambda=g(1)=1$.
The proof of Corollary~\ref{cor:nonpermutation-linearized} gives
$\ker M=\{0\}\cup\mathcal E_{\lambda}$.
Thus Corollary~\ref{cor:coulter-matthews} applies.
Since $n$ is odd, substituting $z_{\lambda,M}=0$ into its boomerang
spectrum formula gives the asserted boomerang spectrum.
It therefore remains to prove that $z_{\lambda,M}=0$.
By Lemma~\ref{lem:z-root-count}, $z_{\lambda,M}$ is twice the number of
solutions of $g(x)+M(x)+\lambda=0$ in $\mathcal D_\lambda\subseteq\F_q$.

We have
\[
 g(x)=x^{(q+9)/2}=\chi(x)x^5\qquad(x\in\F_q).
\]
We show that $g(x)+x^3-x+1$ never vanishes.
Its value at $x=0$ is $1$.
If $x$ is a nonzero square, then
\[
 g(x)+x^3-x+1=x^5+x^3-x+1=(x+1)(x^2+x-1)^2.
\]
Since $n$ is odd, $-1$ is a nonsquare, so $x+1\ne0$ and
$x^2+x-1=(x-1)^2+1\ne0$.
If $x$ is a nonsquare, then
\[
 g(x)+x^3-x+1=-x^5+x^3-x+1
 =(1-x)\bigl((x+1)^4-x\bigr).
\]
Here $x\ne1$, and the second factor cannot vanish because that would
express the nonsquare $x$ as a fourth power.  Hence
$g(x)+M(x)+\lambda$ has no zero in $\F_q$.
Lemma~\ref{lem:z-root-count} therefore gives $z_{\lambda,M}=0$.
\end{proof}

\subsubsection{The linear perturbation \texorpdfstring{$M(x)=x$}{M(x)=x}}
\label{subsec:CM-plusminus-x}

Let $q=3^n$, where $n>1$ is odd, and let $k\ge1$ be an integer
satisfying $\gcd(k,2n)=1$.  Put $g(x)=x^{(3^k+1)/2}$.
Take $M(x)=x$ and $\lambda=1$, and write $z=z_{1,\operatorname{id}}$.
The following corollary gives an explicit gcd formula for $z$.

\begin{corollary}
\label{cor:coulter-matthews-plusminus-x-spectra}
The function
\[
 G(x):=G_{1,\operatorname{id}}(x)
 =x^{(3^k+1)/2}+x+x^{q-1}-1
\]
satisfies
$\delta_{G}=2$, $\beta_{G}=1$, and
\[
 z=\gcd(3^k-2,q-1)+\gcd(3^k-2,q+1)-2.
\]
Its boomerang spectrum is
\begin{equation}\label{eq:coulter-matthews-plusminus-x-spectra}
 \mathcal S_{\BCT}(G)
 =\{\!\{0^{q^2-6q+13+2z},1^{4q-12-2z}\}\!\}.
\end{equation}
In particular, if $k=3$ or $k=n+2$, then $z=0$.
If $k=n-2$, then $z=18$ when $9\mid n$, and $z=0$ otherwise.
\end{corollary}

\begin{proof}
Since $n$ is odd, $\chi(-1)=-1$, so Corollary~\ref{cor:coulter-matthews}
applies with $\lambda=1$ and $M(x)=x$.
Moreover, $\mathcal E_{1}=\{1,-1\}$ by
Lemma~\ref{lem:even-PN-preimages}.
By Lemma~\ref{lem:z-root-count}, $z/2$ counts the roots of
$g(s)+s+1=0$ in $\mathcal D_{1}$.
Since neither $0$ nor $-1$ is a root, we obtain
\begin{equation}\label{eq:coulter-matthews-plusminus-x-roots}
 \frac z2
 =\#\{s\in\F_q\setminus\{1\}:s^{(3^k+1)/2}+s+1=0\}.
\end{equation}

For a root $s$ counted in \eqref{eq:coulter-matthews-plusminus-x-roots},
choose $v\in\F_{q^2}$ with $v^2=s$ and put $u=(v+1)/(v-1)$.
Here $v\ne0,\pm1$, and the root equation gives
$v^{3^k}=-(v^2+1)/v$.  Hence
\[
 u^{3^k}=\frac{v^{3^k}+1}{v^{3^k}-1}
 =\frac{v^2-v+1}{v^2+v+1}
 =\left(\frac{v+1}{v-1}\right)^2=u^2,
\]
so $u^{3^k-2}=1$ and $u\ne1$.
Conversely, if $u\in\F_{q^2}$ satisfies $u^{3^k-2}=1$ and $u\ne1$,
then $u\ne-1$ because $3^k-2$ is odd.  Reversing these equalities shows
that $s=((u+1)/(u-1))^2$ is a root distinct from $1$.
The two choices $v$ and $-v$ correspond to $u$ and $u^{-1}$.
Also, $s\in\F_q$ if and only if $v^q=\pm v$, equivalently
$u^q=u$ or $u^q=u^{-1}$.

Among the solutions of $u^{3^k-2}=1$ in $\F_{q^2}$,
exactly $\gcd(3^k-2,q-1)$ satisfy $u^{q-1}=1$, and exactly
$\gcd(3^k-2,q+1)$ satisfy $u^{q+1}=1$.
Only $u=1$ satisfies both conditions, since $3^k-2$ is odd and
$\gcd(q-1,q+1)=2$.  After excluding $1$, the elements satisfying either condition
form distinct pairs $\{u,u^{-1}\}$, each giving one root in
\eqref{eq:coulter-matthews-plusminus-x-roots}.
This proves the asserted formula for $z$.
Since $n$ is odd, Corollary~\ref{cor:coulter-matthews} gives
\eqref{eq:coulter-matthews-plusminus-x-spectra}.

If $k=3$, then $3^k-2=25$.  For odd $n$, $3^n\equiv2$ or $3\pmod5$,
so both $\gcd(25,q-1)$ and $\gcd(25,q+1)$ equal $1$.
Thus $z=0$.

If $k=n+2$, then $3^k-2=9q-2$, so
$z=\gcd(7,q-1)+\gcd(11,q+1)-2=0$, since
$q\not\equiv1\pmod7$ and $q\not\equiv-1\pmod{11}$ for odd $n$.

If $k=n-2$, then $9(3^k-2)=q-18$, giving
$z=\gcd(17,q-1)+\gcd(19,q+1)-2$.
For odd $n$, $q\not\equiv1\pmod{17}$, whereas
$q\equiv-1\pmod{19}$ holds exactly when $9\mid n$.
Thus $z=18$ when $9\mid n$, and $z=0$ otherwise.
\end{proof}

The proof of Proposition~\ref{prop:coulter-matthews-unbounded-spectra}
shows that, for every positive integer $N$, one can fix a Coulter--Matthews
exponent $(3^k+1)/2$ so that the values of $z$ at $n=9\cdot19^i$,
$0\le i\le N$, are pairwise distinct. These values give $N+1$ distinct
boomerang-spectrum types in \eqref{eq:coulter-matthews-plusminus-x-spectra},
each occurring for infinitely many $n$.

For each odd positive integer $k$, put
\[
 z(k,n):=\gcd(3^k-2,3^n-1)+\gcd(3^k-2,3^n+1)-2
\]
and
\[
 \mathcal Z_k:=
 \{z(k,n):n>1\text{ is odd and }\gcd(k,n)=1\}.
\]

\begin{proposition}[Unbounded number of boomerang-spectrum types for $M(x)=x$]
\label{prop:coulter-matthews-unbounded-spectra}
The set $\mathcal Z_k$ is finite
for every fixed odd positive integer $k$, but the cardinalities
$|\mathcal Z_k|$ are unbounded as $k$ varies.  More precisely, for every
positive integer $N$, there exists an odd $k$ for which $\mathcal Z_k$
contains more than $N$ distinct values greater than $2$.  Each of these
values is attained for infinitely many odd $n>1$ satisfying
$\gcd(k,n)=1$.
\end{proposition}

\begin{proof}
For fixed $k$, both gcd terms defining $z(k,n)$ divide $3^k-2$,
so $\mathcal Z_k$ is finite.

Let $N$ be a positive integer and put $m=N+1$.
Both $2$ and $3$ are primitive modulo $19^2$, and hence modulo every
power of $19$ by the primitive-root lifting theorem; see the proof of
\cite[Ch.~4, Theorem~2, p.~43]{IrelandRosen1990}.
Thus we may choose a positive integer $k$ relatively prime to
$18\cdot19^{m-1}$ such that $3^k\equiv2\pmod{19^m}$.
Put $A=3^k-2$ and $n_j=9\cdot19^j$ for $0\le j<m$.
These $n_j$ are odd and satisfy $\gcd(k,n_j)=1$.
Since $3$ has order $18\cdot19^{r-1}$ modulo $19^r$,
\[
 19^{j+1}\mid3^{n_j}+1,\qquad
 19^{j+2}\nmid3^{n_j}+1.
\]
Write $a_j=\gcd(A,3^{n_j}-1)$ and $b_j=\gcd(A,3^{n_j}+1)$.
Since $n_{j+1}=19n_j$, we have $a_j\mid a_{j+1}$ and
$b_j\mid b_{j+1}$.
Moreover, the exact power of $19$ dividing $b_j$ is $19^{j+1}$,
because $19^m\mid A$.  Consequently,
\[
 18\le z(k,n_0)<z(k,n_1)<\cdots<z(k,n_{m-1}),
\]
giving $m=N+1$ distinct values greater than $2$.

Finally, put $T=\operatorname{ord}_A(3)$.
For every integer $\ell\ge0$, the integer $n_j+2kT\ell$ is odd,
remains relatively prime to $k$, and gives the same two gcd terms as $n_j$.
Hence each displayed value occurs for infinitely many odd $n>1$
satisfying $\gcd(k,n)=1$.
\end{proof}

\subsubsection{Linearized permutation binomials \texorpdfstring{$M_t$}{Mt}}
\label{subsec:CM-binomial}

Let $q=3^n$, where $n>1$ is odd, and let $k\ge1$ be an integer
satisfying $\gcd(k,2n)=1$.  Put $g(x)=x^{(3^k+1)/2}$.
For $t\in\F_q^*$, define
\[
 M_t(x):=t^{(3^k-1)/2}x+t^{-(3^k-1)/2}x^{3^k}.
\]

The identity
\[
                         M_t(ty)=g(t)(y+y^{3^k})
\]
shows that $M_t$ is a permutation if and only if $y\mapsto y+y^{3^k}$ is,
where any nonzero element $y$ of the kernel of the latter map would satisfy
$y^{3^k-1}=-1$.  Since $\gcd(3^k-1,q-1)=2$ and $-1$ is a nonsquare,
this equation has no solution.  Hence $M_t$ is a linearized permutation.

\begin{corollary}
\label{cor:coulter-matthews-binomial-spectra}
For every $t\in\F_q^*$, the polynomial $M_t$ is a linearized permutation.
For every $\lambda\in\F_q^*$ with $\chi(\lambda)=1$, the equation
$g(x)+M_t(x)+\lambda=0$ has exactly one solution in $\F_q$, and
\begin{equation}\label{eq:coulter-matthews-binomial-z}
 z_{\lambda,M_t}
 =\begin{cases}
 0,&\lambda=g(t),\\
 2,&\lambda\ne g(t).
 \end{cases}
\end{equation}
The function $G_{\lambda,M_t}$ is APN and has boomerang
uniformity one.  Its boomerang spectrum is
\[
 \mathcal S_{\BCT}(G_{\lambda,M_t})
 =\begin{cases}
 \{\!\{0^{q^2-6q+13},1^{4q-12}\}\!\},&\lambda=g(t),\\[1mm]
 \{\!\{0^{q^2-6q+17},1^{4q-16}\}\!\},&\lambda\ne g(t).
 \end{cases}
\]
For each fixed $t$, among the nonzero squares $\lambda\in\F_q$,
the first spectrum occurs only when $\lambda=g(t)$, and the second
occurs for the remaining $(q-3)/2$ values.
\end{corollary}

\begin{proof}
Since $n$ is odd and $\lambda$ is a nonzero square,
we have $\chi(-1)=-1$ and $\lambda\in g(\F_q^*)$.
Equivalence~\eqref{eq:lambda-image-equivalence} therefore gives
$-\lambda\notin g(\F_q)$.
First take $t=1$, so $M_1(x)=x+x^{3^k}$ is a linearized permutation.
Since $k$ is odd, $(3^k+1)/2$ is even, and
\[
\begin{aligned}
 g(y^2)+M_1(y^2)&=(y^{3^k}-y)^2,\\
 g(-y^2)+M_1(-y^2)&=-(y^{3^k}+y)^2.
\end{aligned}
\]
The first identity excludes square solutions of
$g(x)+M_1(x)+\lambda=0$, since $-\lambda$ is a nonsquare.
Every nonsquare $x$ has the form $x=-y^2$ with exactly two choices
$y,-y$.  By the second identity,
$g(-y^2)+M_1(-y^2)+\lambda=0$ is equivalent to
$M_1(y)^2=\lambda$.  Since $M_1$ is an odd permutation, its two
solutions are negatives of one another and give exactly one value
of $x$.  Thus $g(x)+M_1(x)+\lambda=0$ has exactly one solution
$s\in\F_q$.

Since $p=3$, \eqref{eq:onepoint-direction-sets} gives
$\mathcal E_{\lambda}=\{u\in\F_q^*:g(u)=\lambda\}$.
If $s\in\mathcal E_{\lambda}$, then $g(s)=\lambda$, and
$g(s)+M_1(s)+\lambda=0$ gives $M_1(s)=-2\lambda=\lambda$.
Hence
\[
 0=M_1(s)^2-g(s)^2=(s^{3^k}-s)^2.
\]
Hence $s^{3^k}=s$, so $s\in\F_3^*$ because $\gcd(k,n)=1$.
Substitution into $M_1(s)=g(s)$ gives $s=-1$, and hence $\lambda=g(-1)=1$.
Conversely, when $\lambda=1$, $s=-1$ satisfies
$g(s)+M_1(s)+\lambda=0$ and belongs to $\mathcal E_{\lambda}$.
By Lemma~\ref{lem:z-root-count}, $z_{\lambda,M_1}$ is twice
the number of roots in $\mathcal D_{\lambda}$, so it equals $0$
when $\lambda=1$ and $2$ otherwise.

For general $t$, scaling gives
\[
 g(ty)+M_t(ty)+\lambda
 =g(t)\left(g(y)+M_1(y)+\frac{\lambda}{g(t)}\right).
\]
Since $\lambda$ and $g(t)$ are nonzero squares, $\lambda/g(t)$ is
also a nonzero square, so the argument for $t=1$ applies.
Also, $ty\in\mathcal E_{\lambda}$ if and only if
$y\in\mathcal E_{\lambda/g(t)}$.
Thus $g(x)+M_t(x)+\lambda=0$ has exactly one solution, which belongs to
$\mathcal E_{\lambda}$ exactly when $\lambda/g(t)=1$.
This proves \eqref{eq:coulter-matthews-binomial-z}.
Corollary~\ref{cor:coulter-matthews} gives the asserted uniformities
and boomerang spectra after substituting the two values of
$z_{\lambda,M_t}$.
Finally, $\lambda=g(t)$ gives the first spectrum, and the remaining
$(q-3)/2$ nonzero squares give the second.
\end{proof}

\subsection[APN functions with boomerang uniformity one from DO PN functions]{APN functions with boomerang uniformity one\\from DO PN functions}
\label{sec:DO-examples}

Recall that a DO PN function is a PN function represented by a
Dembowski--Ostrom polynomial; every such polynomial is even and vanishes
at zero.  Let $q=p^n>3$, where $p$ is an odd prime and $n\ge1$.
For the DO results in this section, we assume
\begin{equation}\label{eq:DO-parameters}
\begin{gathered}
 g:\F_q\longrightarrow\F_q\text{ is a DO PN function},\\
 \lambda\in\F_q^*,\qquad -\lambda\notin g(\F_q).
\end{gathered}
\end{equation}
We determine the boomerang spectra of the APN functions obtained by
taking $M=L_c$, $c\ne0$.

Table~\ref{tab:general-PN-families} collects eight known PN families over
$\F_q$, where $q=p^n$ and $p$ is odd, together with sufficient conditions
for the PN property.  Every function in the table is even and vanishes
at zero.  All families other than $f_2$ are represented by DO polynomials;
the Coulter--Matthews family $f_2$, which is not DO in general, was treated
in Section~\ref{sec:coulter-matthews}.
The families $f_1$, $f_3$, and $f_4$ are
the Gold, Ding--Yuan, and Zha--Kyureghyan--Wang families, respectively;
$f_5$ and $f_6$ are the Bierbrauer families, and $f_7$ and $f_8$ are
the Budaghyan--Helleseth families.
All exponent parameters are integers, with $s>0$ and $k,t\ge0$,
subject to the additional conditions in each row.

\begingroup
\small
\setlength{\tabcolsep}{4pt}
\renewcommand{\arraystretch}{1.2}
\setlength{\LTcapwidth}{\textwidth}
\begin{longtable}{@{}>{\centering\arraybackslash}p{0.07\textwidth}
 >{\raggedright\arraybackslash}p{0.395\textwidth}
 >{\raggedright\arraybackslash}p{0.48\textwidth}@{}}
\caption{Known PN families over $\F_q$, $q=p^n$, in odd characteristic,
with sufficient conditions for the PN property.}\label{tab:general-PN-families}\\
\toprule
Family & PN function $f_i(x)$ & Conditions and sources \\
\midrule
\endfirsthead
\multicolumn{3}{c}{\small Table~\thetable\ (continued)}\\[3pt]
\toprule
Family & PN function $f_i(x)$ & Conditions and sources \\
\midrule
\endhead
\multicolumn{3}{r}{\small Continued on the next page}\\
\endfoot
\bottomrule
\endlastfoot
$f_1$ & $x^{p^k+1}$
 & $k\ge0$, $n/\gcd(n,k)$ odd;\newline
   \cite{DembowskiOstrom1968};
   \cite[Theorem~3.3]{CoulterMatthews1997}. \\
\midrule[0.3pt]
$f_2$ & $x^{(p^k+1)/2}$
 & $p=3$, $k$ odd, $\gcd(n,k)=1$;\newline
   \cite[Theorem~4.1]{CoulterMatthews1997}. \\
\midrule[0.3pt]
$f_3$ & $x^{10}-a x^6-a^2x^2$
 & $p=3$, $n$ odd, $a\in\F_q^*$;\newline
   \cite[Proposition~2.1]{DingYuan2006}. \\
\midrule[0.3pt]
$f_4$ & $x^{p^k+1}-\omega^{p^s-1}x^{p^s+p^{2s+k}}$
 & $n=3s$, $3\nmid s$, $s/\gcd(s,k)$ odd,\newline
   $k\equiv s\pmod3$, $\operatorname{ord}(\omega)=p^n-1$;\newline
   \cite{ZhaKyureghyanWang2009}. \\
\midrule[0.3pt]
$f_5$ & $x^{p^t+1}-\mu x^{p^{2s}+p^{s+t}}$
 & $n=3s$, $s/\gcd(s,t)$ odd,\newline
   $\operatorname{ord}(\mu)=p^{2s}+p^s+1$, and either\newline
   $(s+t)/\gcd(s,t)\equiv0\pmod3$\newline
   or $p^s\equiv p^t\equiv1\pmod3$;\newline
   \cite[Theorem~4]{Bierbrauer2010}. \\
\midrule[0.3pt]
$f_6$ & $x^{p^t+1}-\beta x^{p^{3s}+p^{s+t}}$
 & $n=4s$, $2s/\gcd(2s,t)$ odd,\newline
   $p^s\equiv p^t\equiv1\pmod4$,\newline
   $\operatorname{ord}(\beta)=p^{3s}+p^{2s}+p^s+1$;\newline
   \cite[Theorem~5]{Bierbrauer2010}. \\
\midrule[0.3pt]
$f_7$ & $\displaystyle\begin{aligned}[t]
 &(\beta x)^{p^k+1}-\bigl((\beta x)^{p^k+1}\bigr)^{p^s}\\
 &\qquad+\sum_{i=0}^{s-1}c_i x^{p^i(p^s+1)}
 \end{aligned}$
 & $n=2s$, $s,k>0$, $\gcd(s+k,2s)=\gcd(s+k,s)$,\newline
   $\gcd(p^k+1,p^s+1)\ne\gcd\bigl(p^k+1,(p^s+1)/2\bigr)$,\newline
   $\beta\in\F_q^*$, $c_i\in\F_{p^s}$,\newline
   $L(x)=\sum_{i=0}^{s-1}c_i x^{p^i}$ permutes $\F_{p^s}$;\newline
   \cite{BudaghyanHelleseth2008}; \cite[p.~20]{BudaghyanHelleseth2010};
   \cite[Theorem~1]{BudaghyanHelleseth2011}. \\
\midrule[0.3pt]
$f_8$ & $\displaystyle\begin{aligned}[t]
 &\beta x^{p^s+1}+\xi x^{p^k+p^t}\\
 &\quad+\xi^{p^s}x^{p^{k+s}+p^{s+t}}\\
 &\quad+\sum_{i=0}^{s-1}w_i x^{p^i(p^s+1)}
 \end{aligned}$
 & $n=2s$, $s>0$, $k,t\ge0$, $d=|k-t|$,\newline
   $\gcd(s+d,2s)=\gcd(s+d,s)$,\newline
   $\beta\in\F_q\setminus\F_{p^s}$, $\xi\in\F_q^*$ a nonsquare,\newline
   $w_i\in\F_{p^s}$ for $0\le i<s$;\newline
   \cite{BudaghyanHelleseth2008}; \cite[Theorem~2]{BudaghyanHelleseth2011}. \\
\end{longtable}
\endgroup

Let $g_i$ be PN functions selected from the families in
Table~\ref{tab:general-PN-families}, all defined over the same field
$\F_q$, where $q>3$.
Choose $\lambda_i\in\F_q^*$ with
$-\lambda_i\notin g_i(\F_q)$ and linearized permutations $M_i$, and put
\begin{equation}\label{eq:PN-table-APN-family}
 G_i(x):=g_i(x)+M_i(x)+\lambda_i(x^{q-1}-1).
\end{equation}
Theorem~\ref{thm:even-PN-onepoint} gives
$\delta_{G_i}=2$ and $\beta_{G_i}=1$.
In particular, one may take $M_i(x)=x$ for every row.
For the DO families, Proposition~\ref{prop:derivative-z} and
Corollary~\ref{cor:translated-derivative-family} determine the boomerang
spectra for the choice $M=L_c$ with $c\ne0$.

When $p=3$ and $n>1$ is odd, the families $f_1,f_3,f_4,f_5$
in Table~\ref{tab:general-PN-families}, under the conditions in their
respective rows, also give APN functions with boomerang uniformity one
for the nonpermutation choice $M(x)=x^3-x$ and $\lambda=g(1)$.
Indeed, this follows from Corollary~\ref{cor:nonpermutation-linearized} by taking $c=u=1$.
For each such PN function $g$, we have $\lambda=g(1)\ne0$ and
$-\lambda\notin g(\F_q)$ by Lemma~\ref{lem:even-PN-preimages} and
\eqref{eq:lambda-image-equivalence}.
The $f_4$ family is linearly equivalent to corresponding members of
$f_5$, as explained in~\cite[Section~4.2]{KooKwonKoKim2026}.

We now consider the linearized perturbations $M=L_c$ for a DO PN
function $g$.  The DO representation gives that $L_a$ is
$\F_p$-linear, $L_a(a)=2g(a)$, and $g(tx)=t^2g(x)$ for $t\in\F_p$.
Since $g$ is PN, $L_c$ is a linearized permutation for every $c\ne0$.
Thus we may take $M=L_c$ in Theorem~\ref{thm:even-PN-onepoint}.
This specialization uses the DO hypothesis, unlike the general results of
Section~\ref{sec:onepoint-construction}.  The identity
\[
 g(s)+L_c(s)=g(s+c)-g(c)
\]
reduces the evaluation of $z_{\lambda,L_c}$ to the preimage counts of $g$.

Under \eqref{eq:DO-parameters}, using the count $z_{\lambda,M}$ defined in
\eqref{eq:onepoint-bct-counts}, set
\[
 z_c:=z_{\lambda,L_c}\quad(c\in\F_q^*),\qquad
 A_j:=\#\{c\in\F_q^*:z_c=j\}\quad(j\in\{0,2,4\}).
\]
Under these assumptions, Lemmas~\ref{lem:character-preimage-counts}(ii)
and~\ref{lem:z-root-count} give
\[
 |g^{-1}(\lambda)|=1-\chi(-1),\qquad
 e_{\lambda}=1-\chi(2),\qquad
 d_{\lambda}=q-2+\chi(2).
\]
Hence these three quantities depend only on $q$.

\begin{proposition}[Exact evaluation and distribution of $z_c$]
\label{prop:derivative-z}
For every $c\in\F_q^*$,
\[
 z_c=2\bigl|g^{-1}(g(c)-\lambda)\bigr|
       -2\cdot\mathbf1_{\{8g(c)=-\lambda\}}\in\{0,2,4\}.
\]
The values of $A_j$ are as follows:
\[
\begin{tikzpicture}[baseline=(freq.center)]
\matrix (freq) [matrix of math nodes,
 nodes={anchor=center,minimum width=20mm,inner xsep=5pt,inner ysep=2pt,
        text height=1.8ex,text depth=0.5ex},
 column 1/.style={nodes={minimum width=14mm}},
 column sep=0pt,row sep=0pt,inner sep=0pt] {
 q\bmod8&1&3\ (p\ne3)&3\ (p=3)&5&7\\
 A_0&(q-1)/2&(q-3)/2&(q+1)/2&(q-1)/2&(q-3)/2\\
 A_2&0&4&0&2&2\\
 A_4&(q-1)/2&(q-7)/2&(q-3)/2&(q-5)/2&(q-3)/2\\
};
\draw[line width=0.4pt]
 (freq-1-1.south west) -- (freq-1-6.south east);
\draw[line width=0.4pt]
 (freq-1-1.north east) -- (freq-1-1.east |- freq-4-6.south);
\foreach \col in {2,3,4,5}
 \draw[line width=0.4pt,densely dotted,line cap=round]
 (freq-1-\col.north east) -- (freq-1-\col.east |- freq-4-6.south);
\end{tikzpicture}
\]
\end{proposition}

\begin{proof}
The map $L_c$ is a linearized permutation polynomial.
Lemma~\ref{lem:z-root-count} gives
\[
 \frac{z_c}{2}
 =\#\{s\in\mathcal D_{\lambda}:g(s)+L_c(s)=-\lambda\}
 =\#\{s\in\mathcal D_{\lambda}:g(s+c)=g(c)-\lambda\}.
\]
Before imposing the restriction $s\in\mathcal D_{\lambda}$, the equation
\[
 g(s+c)=g(c)-\lambda,\qquad s\in\F_q,
\]
has $\bigl|g^{-1}(g(c)-\lambda)\bigr|$ solutions.  None of these solutions is $s=0$,
since $\lambda\ne0$.  If a solution lies in
$\mathcal E_{\lambda}$, then $g(s)=-\lambda/2$ and thus
\[
 L_s(c)=L_c(s)=-\lambda-g(s)=g(s)=L_s(s/2).
\]
Since $L_s$ is a permutation, $s=2c$.  Conversely, $s=2c$
is such an excluded solution exactly when $2g(2c)=8g(c)=-\lambda$.
Exactly one solution of $g(s+c)=g(c)-\lambda$ is removed in that case,
proving the general formula for $z_c$.

To determine $A_0,A_2,A_4$, note that \eqref{eq:PN-difference-count},
applied with $t=\lambda$, shows that $g(c)-g(x)=\lambda$ has exactly
$q-1$ ordered solutions $(c,x)\in\F_q^2$.
None has $c=0$, since $-\lambda\notin g(\F_q)$.
For each fixed $c\ne0$, the number of corresponding solutions $x$ is
$\bigl|g^{-1}(g(c)-\lambda)\bigr|$.  Consequently,
\[
 \sum_{c\in\F_q^*}\bigl|g^{-1}(g(c)-\lambda)\bigr|=q-1.
\]
By Lemma~\ref{lem:even-PN-preimages}, exactly $|g^{-1}(\lambda)|$
summands equal $1$, and all remaining summands are $0$ or $2$.
Since there are $q-1$ summands, the numbers of summands equal to $0$
and to $2$ are equal.
Hence, as $c$ ranges over $\F_q^*$,
$2\bigl|g^{-1}(g(c)-\lambda)\bigr|$ takes each of the values $0$ and $4$ exactly
$(q-1-|g^{-1}(\lambda)|)/2$ times, and takes the value $2$ exactly
$|g^{-1}(\lambda)|$ times.

The correction condition $8g(c)=-\lambda$ is equivalent to
$2c\in\mathcal E_{\lambda}$, so it holds for exactly $e_{\lambda}$
nonzero values of $c$.  If $p\ne3$, then under this condition
$g(c)-\lambda=9g(c)=g(3c)\ne0$, so each correction changes a value
$4$ into $2$.  If $p=3$, the correction condition is $g(c)=\lambda$,
so $e_{\lambda}=|g^{-1}(\lambda)|$ and every occurrence of $2$ is
changed to $0$.  Thus $A_0,A_2,A_4$ are given by
\[
\begin{array}{c|ccc}
 & A_0 & A_2 & A_4\\ \hline
 \rule{0pt}{3ex}p\ne3&
 \dfrac{q-1-|g^{-1}(\lambda)|}{2}&
 |g^{-1}(\lambda)|+e_{\lambda}&
 \dfrac{q-1-|g^{-1}(\lambda)|-2e_{\lambda}}{2}\\[2mm]
 p=3&
 \dfrac{q-1+e_{\lambda}}{2}&
 0&
 \dfrac{q-1-e_{\lambda}}{2}
\end{array}
\]
Substituting $|g^{-1}(\lambda)|=1-\chi(-1)$ and
$e_{\lambda}=1-\chi(2)$ from
Lemmas~\ref{lem:character-preimage-counts}(ii) and~\ref{lem:z-root-count},
together with $\chi(-1)=(-1)^{(q-1)/2}$ and
$\chi(2)=(-1)^{(q^2-1)/8}$, gives the stated values of $A_0,A_2,A_4$.
\end{proof}

\Needspace{5\baselineskip}
\begin{corollary}[One-point modifications of translated DO PN functions]
\label{cor:translated-derivative-family}
Let $g:\F_q\to\F_q$ be a DO PN function, where $q>3$ is an odd prime
power, and let $\lambda\in\F_q^*$ satisfy $-\lambda\notin g(\F_q)$.
For every $c\in\F_q^*$, the function
\[
 g(x+c)+\lambda(x^{q-1}-1)
\]
is APN with boomerang uniformity one.  Its DDT and BCT are identical
to those of $G_{\lambda,L_c}$.  Consequently, with $z_c$ as in
Proposition~\ref{prop:derivative-z},
\[
 \mathcal S_{\BCT}(G_{\lambda,L_c})
 =\{\!\{0^{q^2-6q+9-4\chi(2)+2z_c},
          1^{4q-8+4\chi(2)-2z_c}\}\!\}.
\]
For each $j\in\{0,2,4\}$, exactly $A_j$ choices $c\in\F_q^*$
give the BCT spectrum obtained by setting $z_c=j$ in the displayed
BCT formula.
\end{corollary}

\begin{proof}
Since $D_cg(x)=L_c(x)+g(c)$, we have
\[
 g(x+c)+\lambda(x^{q-1}-1)=G_{\lambda,L_c}(x)+g(c).
\]
The constant $g(c)$ cancels from derivatives and output differences.
The assertions therefore follow from Theorem~\ref{thm:even-PN-onepoint},
Corollary~\ref{cor:even-PN-global}, and Proposition~\ref{prop:derivative-z}.
\end{proof}

\begin{remark}
\label{rem:derivative-spectrum-occurrence}
Under \eqref{eq:DO-parameters}, Proposition~\ref{prop:derivative-z}
shows that, for $p\ne3$ and $q>5$, all three numbers $A_0,A_2,A_4$ are positive
if and only if $q\not\equiv1\pmod8$; equivalently, all three BCT
spectra in Corollary~\ref{cor:translated-derivative-family} occur if
and only if $q\not\equiv1\pmod8$.  The exceptional field $q=5$ has
\[
                         (A_0,A_2,A_4)=(2,2,0),
\]
so only two of the three spectra occur.  Suppose now that $p=3$.
Then $A_2=0$ and $A_0,A_4>0$; hence exactly the two spectra
corresponding to $z_c=0$ and $z_c=4$ occur.
\end{remark}

\section{CCZ-inequivalence of the constructed APN functions}
\label{sec:ccz-rigidity}

For each fixed field $\F_q$, all APN functions in our construction share
the differential spectrum given in Corollary~\ref{cor:even-PN-global}(i).
We show that this common spectrum can be shared by many pairwise
CCZ-inequivalent APN functions with boomerang uniformity one.

We now characterize CCZ-equivalence between members of the one-point
construction for arbitrary linearized perturbations.  In particular,
linearly inequivalent even PN functions give rise to CCZ-inequivalent
APN functions under the construction of Theorem~\ref{thm:even-PN-onepoint}.
We use the graph-intersection identity~\eqref{eq:graph-intersection}
from Section~\ref{sec:preliminaries}.
We first show that if $F:\F_q\to\F_q$ differs from a PN function at
exactly $m$ inputs and $m(m+1)<q-q/p$, then every function CCZ-equivalent
to $F$ is EA-equivalent to $F$
(Lemmas~\ref{lem:vertical-kernel} and~\ref{lem:ccz-ea-onepoint}).
Section~\ref{subsec:one-point-rigidity} establishes the equivalence
criterion, Section~\ref{subsec:PN-inequivalent-families} constructs
pairwise CCZ-inequivalent examples from known PN families, and
Section~\ref{subsec:one-point-distance} gives further applications.
When a second even PN function $h$ with $h(0)=0$ is under consideration, for
$\mu\in\F_q^*$ and an $\F_p$-linearized polynomial $N$, we use the
analogous notation
\begin{equation}\label{eq:comparison-family}
 H_{\mu,N}(x):=h(x)+N(x)+\mu(x^{q-1}-1).
\end{equation}
Subtracting the linearized polynomials $M$ and $N$ gives
\begin{equation}\label{eq:linear-perturbation-EA}
 G_{\lambda,M}\sim_{\EA}G_{\lambda,0},\qquad
 H_{\mu,N}\sim_{\EA}H_{\mu,0}.
\end{equation}

\subsection{Characterization of CCZ-equivalence}
\label{subsec:one-point-rigidity}

Let
\begin{equation}\label{eq:vertical-projection}
 \pi:\F_q\times\F_q\longrightarrow\F_q,\qquad \pi(a,b)=a,
 \qquad
 \mathcal K:=\ker\pi=\{0\}\times\F_q.
\end{equation}
For every function $F:\F_q\to\F_q$ and every
$(0,b)\in\mathcal K\setminus\{(0,0)\}$, one has
$\delta_F(0,b)=0$.

\begin{lemma}
\label{lem:vertical-kernel}
Let $q=p^n$, where $p$ is an odd prime and $n\ge1$.
Let $f:\F_q\to\F_q$ be PN, let $F:\F_q\to\F_q$ be any function,
and put $m=\#\{x\in\F_q:F(x)\ne f(x)\}$.
Assume that $m(m+1)<q-q/p$.  Then $\mathcal K$ is the unique
$n$-dimensional $\F_p$-subspace $U\le\F_q\times\F_q$ satisfying
\begin{equation}\label{eq:kernel-zero-subspace}
 U\setminus\{(0,0)\}
 \subseteq
 \{v\in\F_q\times\F_q:\delta_F(v)=0\}.
\end{equation}
\end{lemma}

\begin{proof}
The preceding observation shows that $\mathcal K$ has the required
property.  Let $U$ be an $n$-dimensional $\F_p$-subspace satisfying
\eqref{eq:kernel-zero-subspace}.
Two distinct points $P,Q\in\mathcal G_F$ cannot lie in the same coset
of $U$.  Indeed, put $v=Q-P\ne(0,0)$.  Then
$Q\in\mathcal G_F\cap(\mathcal G_F+v)$, so
$\delta_F(v)\ge1$ by \eqref{eq:graph-intersection}.
Therefore $v\notin U$ by \eqref{eq:kernel-zero-subspace}.
There are $q$ cosets of $U$ and $q$ points in $\mathcal G_F$,
so each coset contains exactly one point of $\mathcal G_F$.

For the cosets $V_1,\ldots,V_q$ of $U$, put
\[
 N_j:=|\mathcal G_f\cap V_j|,\qquad
 m_j:=|(\mathcal G_f\setminus\mathcal G_F)\cap V_j|.
\]
Let $P_j$ be the unique point of $\mathcal G_F\cap V_j$.  Then
\[
 m_j=
 \begin{cases}
  N_j-1,&P_j\in\mathcal G_f,\\
  N_j,&P_j\notin\mathcal G_f.
 \end{cases}
\]
Consequently, $N_j\le1+m_j$.  Since the cosets $V_j$ partition
$\F_q^2$, we also have
$\sum_jm_j=|\mathcal G_f\setminus\mathcal G_F|=m$.
Since $f$ is PN, $\delta_f(v)=1$ for $v\notin\mathcal K$, whereas
$\delta_f(v)=0$ for $v\in\mathcal K\setminus\{(0,0)\}$.
For distinct points $P,Q\in\mathcal G_f\cap V_j$, their difference
$v=Q-P$ belongs to $U\setminus\mathcal K$, since distinct graph points
have different first coordinates.  The PN property gives $\delta_f(v)=1$,
so an ordered pair with difference $v$ is unique.
Conversely, for every $v\in U\setminus\mathcal K$,
\eqref{eq:graph-intersection} gives a unique point
$Q\in\mathcal G_f\cap(\mathcal G_f+v)$.
Setting $P=Q-v$, we obtain distinct points $P,Q\in\mathcal G_f$
lying in a unique common coset $V_j$.
Thus $(P,Q)\mapsto Q-P$ is a bijection between the ordered pairs of
distinct graph points lying in a common coset of $U$ and the elements
of $U\setminus\mathcal K$.
Each coset $V_j$ contains $N_j(N_j-1)$ such ordered pairs.
Counting these pairs and using the preceding bounds gives
\begin{equation}\label{eq:PN-coset-pair-bound}
\begin{aligned}
 q-|U\cap\mathcal K|
   &=\sum_{v\in U\setminus\{(0,0)\}}\delta_f(v)
     =\sum_{j=1}^qN_j(N_j-1)\\
   &\le\sum_{j=1}^qm_j(m_j+1)\le m(m+1).
\end{aligned}
\end{equation}
If $U\ne\mathcal K$, then
$\dim_{\F_p}(U\cap\mathcal K)\le n-1$, so
$|U\cap\mathcal K|\le q/p$.  Equation~\eqref{eq:PN-coset-pair-bound}
would then imply $q-q/p\le m(m+1)$, contrary to the hypothesis.
Thus $U=\mathcal K$.
\end{proof}

\begin{lemma}
\label{lem:ccz-ea-onepoint}
Let $q,f,F,m$ satisfy the hypotheses of Lemma~\ref{lem:vertical-kernel},
and let $H:\F_q\to\F_q$ be any function.
Suppose that $F\sim_{\CCZ}H$, and let $\mathcal A:\F_q^2\to\F_q^2$
be any $\F_p$-affine permutation satisfying
$\mathcal A(\mathcal G_F)=\mathcal G_H$.  Let $\mathcal M$ be the linear
part of $\mathcal A$.  Then, for all $x,y\in\F_q$,
\begin{equation}\label{eq:ccz-ea-normal-form}
\begin{aligned}
 \mathcal M(x,y)&=(M_1(x),M_2(y)+B(x)),\\
 \mathcal A(x,y)&=\mathcal M(x,y)+(a_0,b_0)\\
   &=(M_1(x)+a_0,M_2(y)+B(x)+b_0),
\end{aligned}
\end{equation}
where $M_1,M_2$ are invertible $\F_p$-linear maps, $B$ is
$\F_p$-linear, and $a_0,b_0\in\F_q$.
For this map, the condition $\mathcal A(\mathcal G_F)=\mathcal G_H$
is equivalent to
\begin{equation}\label{eq:ccz-ea-function-identity}
 H(M_1(x)+a_0)=M_2(F(x))+B(x)+b_0\qquad(x\in\F_q).
\end{equation}
Thus $F\sim_{\EA}H$, with the EA-equivalence realized by $\mathcal A$.
\end{lemma}

\begin{proof}
Write $\mathcal A(v)=\mathcal M(v)+(a_0,b_0)$.
For every $v\in\mathcal M^{-1}(\mathcal K)\setminus\{(0,0)\}$,
Lemma~\ref{lem:ddt-entry-transport} gives
$\delta_F(v)=\delta_H(\mathcal M(v))=0$.
Thus the $n$-dimensional subspace $\mathcal M^{-1}(\mathcal K)$
satisfies \eqref{eq:kernel-zero-subspace}.  Lemma~\ref{lem:vertical-kernel}
therefore yields $\mathcal M^{-1}(\mathcal K)=\mathcal K$, and hence
\[
                              \mathcal M(\mathcal K)=\mathcal K.
\]
For $x,y\in\F_q$, linearity gives
$\mathcal M(x,y)=\mathcal M(x,0)+\mathcal M(0,y)$.
Using the projection $\pi$ in \eqref{eq:vertical-projection}, define
$M_1(x):=\pi(\mathcal M(x,0))$, and let $B(x)$ be the second
coordinate of $\mathcal M(x,0)$.  Thus
\[
 \mathcal M(x,0)=(M_1(x),B(x)).
\]
The $\F_p$-linearity of $M_1$ and $B$ follows from that of $\mathcal M$.
Since $\mathcal M(\mathcal K)=\mathcal K$, there is an $\F_p$-linear
map $M_2$ such that
\[
 \mathcal M(0,y)=(0,M_2(y)).
\]
The restriction of $\mathcal M$ to $\mathcal K$ is invertible, so
$M_2$ is invertible.  Moreover,
$\mathcal M^{-1}(\mathcal K)=\mathcal K$; hence $M_1(x)=0$ implies
$x=0$, and $M_1$ is invertible.  This proves
\eqref{eq:ccz-ea-normal-form}.
\end{proof}

Having established the form of every affine CCZ map, we now recover
the underlying PN function.  The next lemma recovers $h+N$ from its one-point
modification $H_{\mu,N}$.  Its conclusion holds among all PN functions.

\begin{lemma}
\label{lem:unique-completion-onepoint}
Let $q=p^n>3$ be odd, let $h:\F_q\to\F_q$ be an even PN function
satisfying $h(0)=0$, let $\mu\in\F_q^*$ satisfy
$-\mu\notin h(\F_q)$, and let $N$ be an $\F_p$-linearized polynomial.
Write $H=H_{\mu,N}$ as in \eqref{eq:comparison-family}.
Suppose that a PN function $f:\F_q\to\F_q$ agrees with the APN function
$H$ at every point except possibly $t\in\F_q$.  Then
\[
                              t=0,\qquad f=h+N.
\]
\end{lemma}

\begin{proof}
Use the notation in
\eqref{eq:core-centered-derivative}--\eqref{eq:onepoint-direction-sets}
with $(g,\lambda)$ replaced by $(h,\mu)$; in particular,
$L_a=L_{a,h}$ and $s_a=L_{a,h}^{-1}(\mu)$.
By Lemma~\ref{lem:even-PN-preimages},
$|\mathcal E_{h,\mu}|\in\{0,2\}$; since $q>3$, the set
$\mathcal D_{h,\mu}$ is nonempty.
Fix $a\in\mathcal D_{h,\mu}$.  Formula \eqref{eq:nonsingleton-fibers}
in the proof of Theorem~\ref{thm:even-PN-onepoint} gives precisely
the following two-element fibers of $D_aH$:
\begin{equation}\label{eq:completion-nonsingleton-fibers}
                         \{0,s_a\},\qquad\{-a,-a-s_a\}.
\end{equation}

In particular, $H$ is not PN, so $f(t)\ne H(t)$.  The derivatives
$D_af$ and $D_aH$ can differ only at $t$ and $t-a$.
The map $D_aH$ takes exactly two output values twice and omits exactly
two output values.  Since $D_af$ is a permutation and differs from
$D_aH$ only at the inputs $t$ and $t-a$, the set $\{t,t-a\}$ must
meet each of the two-element fibers in
\eqref{eq:completion-nonsingleton-fibers} in exactly one point.
Suppose first that $t\in\{0,s_a\}$ and
$t-a\in\{-a,-a-s_a\}$.  Then
\[
 t\in\{0,s_a\}\cap\{0,-s_a\}=\{0\},
\]
since $s_a\ne0$ and the characteristic is odd.
If the memberships are reversed, then $t-a\in\{0,s_a\}$ and
$t\in\{-a,-a-s_a\}$, so
\begin{equation}\label{eq:completion-fiber-intersection}
 t\in\{a,a+s_a\}\cap\{-a,-a-s_a\}.
\end{equation}
Since $a\ne0$ and the characteristic is odd, the intersection in
\eqref{eq:completion-fiber-intersection} can be nonempty only if
$a+s_a=0$ or $a+s_a=-a$.  Hence $s_a=-a$ or $s_a=-2a$.
If $s_a=-a$, then
$\mu=L_a(-a)=-2h(a)$, contradicting $a\in\mathcal D_{h,\mu}$.
If $s_a=-2a$, then
\[
 \mu=L_a(-2a)=h(-a)-h(-2a)-h(a)=-h(2a),
\]
where the last equality uses evenness of $h$.  This contradicts
$-\mu\notin h(\F_q)$.  Thus the reversed memberships are impossible,
and $t=0$.

Now $f=h+N$ outside zero.  For every $a\ne0$,
the values of $D_a(h+N)$ at the inputs $x=0$ and $x=-a$ are
\[
                         N(a)+h(a),\qquad N(a)-h(a),
\]
whereas the values of $D_af$ at these inputs are
\[
                         N(a)+h(a)-f(0),\qquad N(a)-h(a)+f(0).
\]
The two derivatives agree at every other input.  Since both are
permutations, these two unordered pairs of output values must be equal,
which gives
\[
                              f(0)=0\qquad\text{or}\qquad f(0)=2h(a).
\]
If $f(0)\ne0$, then $h(a)=f(0)/2$ for every $a\ne0$.  This is impossible,
because $f(0)/2\ne0$ would then have the $q-1>2$ preimages $a\ne0$,
whereas every nonzero image value of $h$ has exactly two preimages,
by Lemma~\ref{lem:even-PN-preimages}.
Thus $f(0)=0$, and
$f=h+N$.
\end{proof}

\begin{remark}
\label{rem:unique-PN-completion}
The function $f$ in Lemma~\ref{lem:unique-completion-onepoint} is
assumed only to be PN; neither evenness nor $f(0)=0$ is imposed.
Thus the lemma proves uniqueness
within the full class of PN functions.
\end{remark}

\begin{theorem}[Criterion for CCZ-equivalence of one-point modifications]
\label{thm:onepoint-ccz-rigidity}
Let $q=p^n>3$, where $p$ is an odd prime and $n\ge1$, and let
$g,h:\F_q\to\F_q$ be even PN functions satisfying $g(0)=h(0)=0$.
Let $M,N$ be $\F_p$-linearized polynomials, and choose
$\lambda,\mu\in\F_q^*$ such that $-\lambda\notin g(\F_q)$ and
$-\mu\notin h(\F_q)$.
Then $G_{\lambda,M}\sim_{\CCZ}H_{\mu,N}$ if and only if there exist
invertible $\F_p$-linear maps $M_1,M_2:\F_q\to\F_q$ such that
\[
 h(M_1(x))=M_2(g(x))\quad(x\in\F_q),\qquad M_2(\lambda)=\mu.
\]
\end{theorem}

\begin{proof}
By \eqref{eq:linear-perturbation-EA}, it suffices to prove the theorem
when $M=N=0$.  Assume $G_{\lambda,0}\sim_{\CCZ}H_{\mu,0}$.
Since $G_{\lambda,0}$ differs from the PN function $g$ only at zero
and $2<q-q/p$ for every odd $q>3$, Lemma~\ref{lem:ccz-ea-onepoint} implies
$G_{\lambda,0}\sim_{\EA}H_{\mu,0}$.
Therefore, there exists an $\F_p$-affine permutation $\mathcal A$
of the form \eqref{eq:ccz-ea-normal-form} mapping the graph of
$G_{\lambda,0}$ onto that of $H_{\mu,0}$.
Equation~\eqref{eq:ccz-ea-function-identity} then gives
\begin{equation}\label{eq:onepoint-ccz-function-identity}
 H_{\mu,0}(M_1(x)+a_0)
 =M_2(G_{\lambda,0}(x))+B(x)+b_0\qquad(x\in\F_q).
\end{equation}
Define a function $f$ by
\begin{equation}\label{eq:transported-PN-completion}
 f(M_1(x)+a_0)=M_2(g(x))+B(x)+b_0.
\end{equation}
The function $f$ is PN because it is EA-equivalent to $g$.  For $x\ne0$,
one has $g(x)=G_{\lambda,0}(x)$, so
\eqref{eq:onepoint-ccz-function-identity} and
\eqref{eq:transported-PN-completion} show that $f$ agrees with
$H_{\mu,0}$ outside the single point $a_0$.  Lemma
\ref{lem:unique-completion-onepoint} therefore gives
\[
                              a_0=0,\qquad f=h.
\]
Evaluating \eqref{eq:onepoint-ccz-function-identity} and
\eqref{eq:transported-PN-completion} at $x=0$, and using
$g(0)=h(0)=B(0)=0$, gives
\[
 -\mu=M_2(-\lambda)+b_0,\qquad 0=b_0.
\]
Hence $b_0=0$ and $M_2(\lambda)=\mu$.
Equation~\eqref{eq:transported-PN-completion} now yields
\[
 h(M_1(x))=M_2(g(x))+B(x).
\]
Replacing $x$ by $-x$ in the preceding equation, and using the
evenness of $g,h$ and the linearity of $M_1,B$, gives
\[
 h(M_1(x))=M_2(g(x))-B(x).
\]
Comparing the two equations yields $2B(x)=0$ for every $x\in\F_q$.
Since $p$ is odd, $B=0$, and hence
\[
 h(M_1(x))=M_2(g(x)).
\]

Conversely, suppose that invertible $\F_p$-linear maps $M_1,M_2$
satisfy the stated identities.  Define
\[
 \mathcal A(x,y)=(M_1(x),M_2(y)).
\]
This map is invertible because $M_1$ and $M_2$ are invertible.
For $x\ne0$, the identity $h(M_1(x))=M_2(g(x))$ gives
\[
 \mathcal A\bigl(x,G_{\lambda,0}(x)\bigr)
 =\bigl(M_1(x),H_{\mu,0}(M_1(x))\bigr).
\]
The same equality holds at $x=0$ because $M_2(\lambda)=\mu$.
Thus $\mathcal A$ maps the graph of $G_{\lambda,0}$ onto that of
$H_{\mu,0}$, proving $G_{\lambda,0}\sim_{\CCZ}H_{\mu,0}$.
\end{proof}

Recall from Section~\ref{sec:preliminaries} that $g,h:\F_q\to\F_q$
are \emph{linearly equivalent} if there exist invertible
$\F_p$-linear maps $M_1,M_2$ such that
\[
 h(M_1(x))=M_2(g(x))\qquad(x\in\F_q).
\]

\begin{corollary}
\label{cor:inequivalent-cores}
Under the hypotheses of Theorem~\ref{thm:onepoint-ccz-rigidity}, if
$g$ and $h$ are not linearly equivalent, then the APN functions
$G_{\lambda,M}$ and $H_{\mu,N}$ are not CCZ-equivalent.
\end{corollary}

\begin{proof}
If $G_{\lambda,M}$ and $H_{\mu,N}$ were CCZ-equivalent,
Theorem~\ref{thm:onepoint-ccz-rigidity} would imply that $g$ and $h$
are linearly equivalent, a contradiction.
\end{proof}

\subsection{CCZ-inequivalent APN functions from known PN families}
\label{subsec:PN-inequivalent-families}

We now apply Corollary~\ref{cor:inequivalent-cores} to the PN families
in Table~\ref{tab:general-PN-families}.
Let $g_i$ be selected PN functions from these families, all defined over
the same field $\F_q$ with $q>3$.
Choose $\lambda_i\in\F_q^*$ with $-\lambda_i\notin g_i(\F_q)$,
let $M_i$ be arbitrary $\F_p$-linearized polynomials, and define $G_i$
by the formula in \eqref{eq:PN-table-APN-family}.
Theorem~\ref{thm:even-PN-onepoint} shows that each $G_i$ is APN.
If the selected $g_i$ are pairwise linearly inequivalent, then
Corollary~\ref{cor:inequivalent-cores} shows that the $G_i$ are
pairwise CCZ-inequivalent, independently of the choices of
$\lambda_i$ and $M_i$ satisfying these conditions.
In particular, choosing $M_i(x)=x$ gives boomerang uniformity one.

\paragraph{Linear inequivalence of the DO PN functions.}
We use the nucleus orders of associated semifields, as in
\cite[Section~5.3]{KooKwonKoKim2026}, but need only linear
inequivalence of the PN functions.  A DO PN function $g$ defines
a commutative presemifield
$\mathbb P_g=(\F_q,+,\circ_g)$ by
\[
 x\circ_g y:=L_{x,g}(y)=g(x+y)-g(x)-g(y).
\]
Here $L_{x,g}$ denotes the linearized derivative of $g$ with increment
$x$, as defined in Section~\ref{sec:preliminaries}.
An isotopy between two presemifields is given by three invertible
$\F_p$-linear maps $A,B,C$ such that
$C(x\circ y)=A(x)*B(y)$; it is a strong isotopy when $A=B$.
Both isotopy and strong isotopy are equivalence relations.

We define an associated commutative semifield explicitly using the
standard construction in
\cite{AndreoliEtAl2025,CoulterHenderson2008}.
Since $g$ is DO PN, the map $L_{1,g}(x)=1\circ_g x$ is an
$\F_p$-linear permutation.
Define
\[
 \mathbb S_g:=(\F_q,+,*_g),\qquad
 x*_g y:=L_{1,g}^{-1}(x)\circ_g L_{1,g}^{-1}(y).
\]
This multiplication is commutative and distributive and has no zero
divisors.  Since $g(2)=4g(1)$ for a DO polynomial, its multiplicative
identity is
\[
 1_{\mathbb S_g}=L_{1,g}(1)=1\circ_g1
              =g(2)-2g(1)=2g(1)\ne0.
\]
Indeed,
\[
 L_{1,g}(1)*_g x=1\circ_g L_{1,g}^{-1}(x)
           =L_{1,g}\bigl(L_{1,g}^{-1}(x)\bigr)=x,
\]
and commutativity gives the identity on the other side.  Moreover,
\[
 L_{1,g}(x)*_g L_{1,g}(y)=x\circ_g y,
\]
so $(A,B,C)=(L_{1,g},L_{1,g},\operatorname{id})$ gives a strong isotopy
from $\mathbb P_g$ to $\mathbb S_g$.

For a semifield $\mathbb S=(\F_q,+,*)$, its left, middle, and right
nuclei are, respectively,
\[
\begin{aligned}
 N_{\ell}(\mathbb S)
   &:=\{a\in\F_q:(a*x)*y=a*(x*y)\text{ for all }x,y\in\F_q\},\\
 N_m(\mathbb S)
   &:=\{a\in\F_q:(x*a)*y=x*(a*y)\text{ for all }x,y\in\F_q\},\\
 N_r(\mathbb S)
   &:=\{a\in\F_q:(x*y)*a=x*(y*a)\text{ for all }x,y\in\F_q\},
\end{aligned}
\]
and its nucleus is
$N(\mathbb S):=N_{\ell}(\mathbb S)\cap N_m(\mathbb S)\cap N_r(\mathbb S)$.
For a commutative semifield,
$N(\mathbb S)=N_{\ell}(\mathbb S)=N_r(\mathbb S)\subseteq N_m(\mathbb S)$,
and its center equals $N(\mathbb S)$.
The orders of these nuclei are invariant under isotopy
\cite[Section~2.2]{AndreoliEtAl2025}; hence every semifield isotopic
to $\mathbb P_g$ has the same nucleus orders as $\mathbb S_g$.

Suppose that $g,h$ are linearly equivalent DO PN functions, so that
$h(M_1(x))=M_2(g(x))$ for invertible $\F_p$-linear maps $M_1,M_2$.
Taking linearized derivatives gives
\begin{equation}\label{eq:linear-equivalence-strong-isotopy}
 M_1(x)\circ_h M_1(y)
   =L_{M_1(x),h}(M_1(y))
   =M_2(L_{x,g}(y))=M_2(x\circ_g y).
\end{equation}
Thus $\mathbb P_g$ and $\mathbb P_h$ are strongly isotopic.
Since $\mathbb S_g$ is strongly isotopic to $\mathbb P_g$
and $\mathbb S_h$ is strongly isotopic to $\mathbb P_h$, transitivity
of strong isotopy shows that $\mathbb S_g$ and $\mathbb S_h$
are strongly isotopic.  Therefore
\begin{equation}\label{eq:linear-equivalence-nuclei}
 \bigl(|N(\mathbb S_g)|,|N_m(\mathbb S_g)|\bigr)
 =\bigl(|N(\mathbb S_h)|,|N_m(\mathbb S_h)|\bigr).
\end{equation}
Consequently, a difference in either order proves linear inequivalence.
Equality of the two pairs is only a necessary condition for linear
equivalence, not a sufficient one.

\Needspace{13\baselineskip}
\begin{theorem}[Three and six CCZ-inequivalent APN functions]
\label{thm:three-odd-characteristic-classes}
Let $p$ be an odd prime, let $\ell\ge1$ be an integer, and put $q=p^{15\ell}$.
Let $\omega$ be a primitive element of $\F_q$, put
$\mu=\omega^{p^{5\ell}-1}$, and define
\begin{equation}\label{eq:three-PN-inputs}
\begin{aligned}
 g_0(x)&=x^2,\\
 g_1(x)&=x^{p^{5\ell}+1},\\
 g_5(x)&=x^{p^\ell+1}-\mu x^{p^{10\ell}+p^{6\ell}}.
\end{aligned}
\end{equation}
\begin{enumerate}
\item[(i)] The functions $g_0,g_1,g_5$ are pairwise linearly inequivalent
DO PN functions.
\item[(ii)] If $\ell=8r$ for an integer $r\ge1$, put
$\theta=\omega-\omega^{p^{60r}}$ and define
\[
\begin{aligned}
 g_6(x)&=x^{p^{4r}+1}
          -\omega^{p^{30r}-1}x^{p^{90r}+p^{34r}},\\
 g_7(x)&=x^{p^{60r}+1}+x^{p^r+1}
          -x^{p^{61r}+p^{60r}},\\
 g_8(x)&=\theta x^{p^{60r}+1}+\omega x^{p^{3r}+1}
          +\omega^{p^{60r}}x^{p^{63r}+p^{60r}}.
\end{aligned}
\]
Then $g_0,g_1,g_5,g_6,g_7,g_8$ are pairwise linearly inequivalent
DO PN functions over $\F_{p^{120r}}$.
\end{enumerate}
For either collection, choose $\lambda_i\in\F_q^*$ with
$-\lambda_i\notin g_i(\F_q)$ for each index $i$, and define $G_i$ by
\eqref{eq:PN-table-APN-family} with $M_i(x)=x$.
Then the corresponding functions $G_i$ form a family of three pairwise
CCZ-inequivalent APN functions in case~\textup{(i)}, or six in
case~\textup{(ii)}, all with boomerang uniformity one.
\end{theorem}

\begin{proof}
\textup{(i)} The functions $g_0,g_1$ belong to the Gold family $f_1$
in Table~\ref{tab:general-PN-families} and satisfy its PN conditions.
For $g_5$, take $s=5\ell$ and $t=\ell$ in the Bierbrauer family $f_5$.
Then $s/\gcd(s,t)=5$ is odd,
$(s+t)/\gcd(s,t)=6\equiv0\pmod3$, and
$\operatorname{ord}(\mu)=p^{10\ell}+p^{5\ell}+1$.
Thus all three are DO PN functions.

Write $\mathbb S_i:=\mathbb S_{g_i}$ for the semifield constructed above
for each DO PN function $g_i$.
Since $\mathbb S_0$ is isotopic to $\F_q$, both its nucleus orders
equal $q$.
For $g_1$, \cite[Lemma~3.8]{CoulterHenderson2008}, with
$\gcd(15\ell,5\ell)=5\ell$, gives both nucleus orders $p^{5\ell}$.
For $g_5$, \cite[Theorems~4.3 and~A.4]{GologluKolsch2026}, with
$m=5\ell$, $k=\ell$, and $d=\gcd(m,k)=\ell$, give both nucleus
orders $p^\ell$.
In summary,
\begin{equation}\label{eq:three-nucleus-orders}
 \bigl(|N(\mathbb S_i)|,|N_m(\mathbb S_i)|\bigr)
 =\begin{cases}
 (p^{15\ell},p^{15\ell}),&i=0,\\
 (p^{5\ell},p^{5\ell}),&i=1,\\
 (p^\ell,p^\ell),&i=5.
 \end{cases}
\end{equation}
The three pairs are distinct.  By
\eqref{eq:linear-equivalence-nuclei}, the PN functions are pairwise
linearly inequivalent.

\textup{(ii)} Suppose now that $\ell=8r$.
It is straightforward to verify the PN conditions in
Table~\ref{tab:general-PN-families} for the Bierbrauer function $g_6$
from $f_6$ and the Budaghyan--Helleseth functions $g_7,g_8$ from
$f_7,f_8$, respectively.
For $g_8$, take $s=60r$, $k=3r$, $t=0$, $\beta=\theta$,
$\xi=\omega$, and $w_i=0$ for all $i$.
Since $\omega\notin\F_{p^{60r}}$, we have $\theta\ne0$ and
$\theta^{p^{60r}}=-\theta$; in particular,
$\theta\notin\F_{p^{60r}}$.
Moreover, $\omega$ is a nonsquare and
$\gcd(63r,120r)=\gcd(63r,60r)=3r$.
Thus $g_6,g_7,g_8$ are DO PN functions.

For $g_6$, \cite[Theorem~4.5]{MarinoPolverino2012}, with base-field
order $p^{2r}$ and parameters $m=15$, $n=2$, gives nucleus and
middle-nucleus orders $p^{2r}$ and $p^{4r}$, respectively.

The presemifields associated with $g_7$ and $\theta^{-1}g_8$ belong
to the Budaghyan--Helleseth family in
\cite[Section~4.1]{MarinoPolverino2012}, with base-field orders
$p^r$ and $p^{3r}$, exponent pairs $(60,1)$ and $(20,1)$, and
nonsquare parameters $\theta^{-1}$ and $\omega$, respectively.
Here $\theta^{-1}$ is a nonsquare since
$\theta^{p^{60r}}=-\theta$ and $p^{60r}\equiv1\pmod4$.
Since $g_8$ and $\theta^{-1}g_8$ are linearly equivalent, their
associated semifields have the same nucleus orders by
\eqref{eq:linear-equivalence-nuclei}.
Therefore \cite[Theorem~4.1]{MarinoPolverino2012} gives
\[
 \bigl(|N(\mathbb S_i)|,|N_m(\mathbb S_i)|\bigr)
 =\begin{cases}
 (p^{2r},p^{4r}),&i=6,\\
 (p^r,p^{2r}),&i=7,\\
 (p^{3r},p^{6r}),&i=8.
 \end{cases}
\]
Together with the three pairs in \eqref{eq:three-nucleus-orders}
for $\ell=8r$, these give six distinct pairs.  Hence
\eqref{eq:linear-equivalence-nuclei} proves that the six $g_i$
are pairwise linearly inequivalent.

In either part, all $g_i$ are even and vanish at zero.
Theorem~\ref{thm:even-PN-onepoint} provides $(q-1)/2$ choices of
each $\lambda_i$ and gives $\delta_{G_i}=2$ and $\beta_{G_i}=1$.
Corollary~\ref{cor:inequivalent-cores} gives their pairwise
CCZ-inequivalence.
\end{proof}

In characteristic three and odd dimension $15\ell$, the next
result adds two linearly inequivalent Ding--Yuan functions and
non-DO Coulter--Matthews
functions of distinct algebraic degrees to part~(i) of
Theorem~\ref{thm:three-odd-characteristic-classes}.

\Needspace{11\baselineskip}
\begin{corollary}[CCZ-inequivalent APN functions in characteristic three]
\label{cor:ternary-classes}
Let $\ell>1$ be odd, put $n=15\ell$ and $q=3^n$, and let
$g_0,g_1,g_5$ be as in \eqref{eq:three-PN-inputs}, with $p=3$.
Put
\[
 \mathcal I_n:=\{k:1<k<n,\ k\text{ odd},\ \gcd(k,n)=1\},
\]
and define
\[
\begin{aligned}
 g_3(x)&=x^{10}-x^6-x^2,\qquad
 \widetilde g_3(x)=x^{10}+x^6-x^2,\\
 h_k(x)&=x^{(3^k+1)/2}\quad(k\in\mathcal I_n).
\end{aligned}
\]
For $i\in\{0,1,3,5\}$, define $G_i$ by
\eqref{eq:PN-table-APN-family} with $M_i(x)=x$ and
$\lambda_i=g_i(1)$; explicitly,
\[
 \lambda_0=\lambda_1=1,\qquad
 \lambda_3=-1,\qquad \lambda_5=1-\mu.
\]
Also put
\[
\begin{aligned}
 \widetilde G_3(x)&=\widetilde g_3(x)+x+(x^{q-1}-1),\\
 H_k(x)&=h_k(x)+x+(x^{q-1}-1)\quad(k\in\mathcal I_n).
\end{aligned}
\]
Then the $\varphi(n)/2+4$ PN functions $g_0,g_1,g_3,\widetilde g_3,g_5$ and
$h_k$ $(k\in\mathcal I_n)$ are pairwise linearly inequivalent,
where $\varphi$ denotes Euler's totient function.
The corresponding functions $G_0,G_1,G_3,\widetilde G_3,G_5$ and
$H_k$ $(k\in\mathcal I_n)$ are pairwise CCZ-inequivalent APN
functions with boomerang uniformity one.  In particular, if
$n=15\ell$ with $\ell>5$ prime, we obtain $4\ell$ pairwise
CCZ-inequivalent APN functions with boomerang uniformity one
over $\F_{3^n}$.
\end{corollary}

\begin{proof}
The functions $g_3$ and $\widetilde g_3$ are the Ding--Yuan functions
in the $f_3$ row with $a=1$ and $a=-1$, respectively.
Each $h_k$ is PN by the $f_2$ row, since $k$ is
odd and $\gcd(k,n)=1$.
Theorem~\ref{thm:three-odd-characteristic-classes}(i) distinguishes
$g_0,g_1,g_5$ from one another.
The semifields associated with $g_3$ and $\widetilde g_3$ both have
nucleus and middle-nucleus orders $(3,3)$
\cite[Section~3]{AndreoliEtAl2025}.
Since $\ell>1$, this pair differs from each pair in
\eqref{eq:three-nucleus-orders}.  Thus
\eqref{eq:linear-equivalence-nuclei} distinguishes each of
$g_3,\widetilde g_3$ from $g_0,g_1,g_5$.
By \cite[Theorem~4.5]{CoulterHenderson2008}, the presemifields
associated with $g_3$ and $\widetilde g_3$ are not isotopic for
odd $n\ge5$.  Hence
\eqref{eq:linear-equivalence-strong-isotopy} shows that these two
functions are not linearly equivalent.
Thus the five DO PN functions are pairwise linearly inequivalent and have
algebraic degree two.

For every $k\in\mathcal I_n$, the exponent of $h_k$ is less than
$q-1$ and has the ternary expansion
\[
 \frac{3^k+1}{2}=2+\sum_{j=1}^{k-1}3^j.
\]
Hence $\deg_{\mathrm{alg}}h_k=k+1>2$.  These degrees are distinct
as $k$ varies.  Since algebraic degree is preserved by linear
equivalence, the $h_k$ are pairwise linearly inequivalent and
linearly inequivalent to the five DO functions.
Since $n$ is odd, pairing each reduced residue $k$ modulo $n$
with $n-k$ shows that exactly half of the $\varphi(n)$ reduced
residues in $\{1,\ldots,n-1\}$ are odd.  Excluding $k=1$ gives
$|\mathcal I_n|=\varphi(n)/2-1$.  Thus the stated collection has
$\varphi(n)/2+4$ pairwise linearly inequivalent PN functions.

All these PN functions are even and vanish at zero.
Since $q\equiv3\pmod4$, Lemma~\ref{lem:even-PN-preimages} and
\eqref{eq:lambda-image-equivalence} give
$\lambda_i=g_i(1)\ne0$ and $-\lambda_i\notin g_i(\F_q)$.
For $\widetilde g_3$ and each $h_k$, the same argument gives
$\lambda=\widetilde g_3(1)=h_k(1)=1$.
Theorem~\ref{thm:even-PN-onepoint} consequently gives
$\delta_{G_i}=\delta_{\widetilde G_3}=\delta_{H_k}=2$ and~\mbox{$\beta_{G_i}=\beta_{\widetilde G_3}=\beta_{H_k}=1$}.
Therefore, by Corollary~\ref{cor:inequivalent-cores}, we obtain
$\varphi(n)/2+4$ pairwise CCZ-inequivalent APN functions with
boomerang uniformity one.
If $\ell>5$ is prime, then
$\varphi(15\ell)=\varphi(15)\varphi(\ell)=8(\ell-1)$,
so $\varphi(n)/2+4=4\ell$, proving the final assertion.
\end{proof}

In the proof of Corollary~\ref{cor:ternary-classes}, we use algebraic degree
to distinguish the Coulter--Matthews functions $h_k$ from one another and
from the five DO PN functions.
The APN functions $G_i$, $\widetilde G_3$, and $H_k$ all have
maximal algebraic degree $2n$ by
Corollary~\ref{cor:onepoint-maximal-degree} in
Section~\ref{subsec:one-point-distance}.

As illustrated by $g_0,g_1$ in
Theorem~\ref{thm:three-odd-characteristic-classes} and by
$g_3,\widetilde g_3$ and the functions $h_k$ in
Corollary~\ref{cor:ternary-classes}, different parameter
choices within a single family $f_i$ in
Table~\ref{tab:general-PN-families} can yield two or more pairwise
linearly inequivalent PN functions.  By
Corollary~\ref{cor:inequivalent-cores}, our construction then yields
pairwise CCZ-inequivalent APN functions with boomerang uniformity one.

\subsection{Equivalence classes, algebraic degree, and Hamming distance}
\label{subsec:one-point-distance}

We apply Lemma~\ref{lem:ccz-ea-onepoint} to functions obtained by
changing $m$ values of a PN function, where $m(m+1)<q-q/p$.
Several classes for which CCZ-equivalence and EA-equivalence coincide
are known; see Budaghyan's monograph~\cite[Chapter~3]{Budaghyan2014}
for background.
These include planar functions over
$\F_{p^n}$~\cite{BudaghyanHelleseth2008,KyureghyanPott2008},
Boolean functions over $\F_{2^n}$~\cite{BudaghyanCarlet2010}, and
vectorial bent Boolean functions over
$\F_{2^n}$~\cite{BudaghyanCarlet2011}.
The two equivalences also coincide for pairs of quadratic APN
functions over $\F_{2^n}$~\cite{Yoshiara2012} and for pairs of
noninvertible power functions over $\F_{p^n}$
\cite{Dempwolff2018,Dempwolff2022}.
For these last two families, both functions in the comparison are
required to belong to the specified family.

In this section, we provide another family of functions for which
CCZ-equivalence coincides with EA-equivalence.
Corollary~\ref{cor:ccz-classes-near-PN}(ii) shows that the entire CCZ
class of each function satisfying a Hamming-distance bound relative
to a PN function coincides with its EA class.
Corollary~\ref{cor:ccz-classes-near-PN}(i) also shows that the property
of having maximal algebraic degree is preserved under CCZ-equivalence.

Let $q=p^n>2$, where $p$ is a prime and $n\ge1$.
For functions $F,H:\F_q\to\F_q$, their Hamming distance is
\[
 d(F,H):=\#\{x\in\F_q:F(x)\ne H(x)\}.
\]
We use the following notation for equivalence classes among all
functions from $\F_q$ to itself:
\[
\begin{aligned}
 \mathcal C_{\CCZ}(F)&:=\{H:\F_q\to\F_q:H\sim_{\CCZ}F\},\\
 \mathcal C_{\EA}(F)&:=\{H:\F_q\to\F_q:H\sim_{\EA}F\}.
\end{aligned}
\]
For a family $S$ of functions from $\F_q$ to itself, define
\[
 \mathcal C_{\CCZ}(S):=\bigcup_{F\in S}\mathcal C_{\CCZ}(F),
 \qquad
 \mathcal C_{\EA}(S):=\bigcup_{F\in S}\mathcal C_{\EA}(F).
\]
We say that $S$ is invariant under CCZ-equivalence if
$\mathcal C_{\CCZ}(S)=S$.
Similarly, $S$ is invariant under EA-equivalence if
$\mathcal C_{\EA}(S)=S$.
In this section, saying that CCZ-equivalence and EA-equivalence
coincide for a family $S$ means that
\[
 \mathcal C_{\CCZ}(F)=\mathcal C_{\EA}(F)
 \qquad\text{for every }F\in S.
\]
This is stronger, in general, than coincidence for pairs of functions
in $S$, which means that
\[
 F\sim_{\CCZ}H
 \quad\Longleftrightarrow\quad
 F\sim_{\EA}H
 \qquad\text{for every }F,H\in S.
\]
The two conditions are equivalent when $S$ is invariant under
CCZ-equivalence.  However, the invariance condition
$\mathcal C_{\CCZ}(S)=S$ alone does not imply either condition.
Since EA-equivalence implies CCZ-equivalence,
\[
 S\subseteq\mathcal C_{\EA}(S)\subseteq\mathcal C_{\CCZ}(S).
\]
Define the family of functions of maximal algebraic degree by
\[
 \mathcal T:=\{F:\F_q\to\F_q:
                 \deg_{\mathrm{alg}}F=n(p-1)\}.
\]
When $p$ is odd, for each integer $m\ge0$, define
\[
 \mathcal C_m:=\{F:\F_q\to\F_q:
 d(F,f)\le m\text{ for some PN function }f\}.
\]
Thus $\mathcal C_0$ is the set of PN functions,
$\mathcal C_1$ consists of PN functions and their one-point
modifications, and
\[
 \mathcal C_0\subseteq\mathcal C_1\subseteq\cdots.
\]

The following degree criterion holds for every prime $p$, including $p=2$.
Let $q=p^n$ and write $F(x)=\sum_{i=0}^{q-1}c_i x^i$ for a
function $F:\F_q\to\F_q$.
The finite-field Lagrange interpolation formula
\cite[Theorem~2.1.131, p.~26]{MullenPanario2013} gives
$c_{q-1}=-\sum_{x\in\F_q}F(x)$, and~hence
\begin{equation}\label{eq:maximal-degree-sum}
 \deg_{\mathrm{alg}}F=n(p-1)
 \quad\Longleftrightarrow\quad\sum_{x\in\F_q}F(x)\ne0.
\end{equation}
Indeed, $\sum_x1=q=0$ in $\F_q$, while
$\sum_xx^i=0$ for $1\le i<q-1$ and $\sum_xx^{q-1}=-1$.
The degree criterion then follows from the $p$-weight formula in
Section~\ref{sec:preliminaries}.

For nonlinear functions, algebraic degree is EA-invariant, but it is
not CCZ-invariant in general.  Nevertheless, the property of having
maximal algebraic degree is CCZ-invariant, as shown in part~(i) below.

Part~(ii) also presents a new class $\mathcal C_m$ of functions for
which CCZ-equivalence coincides with EA-equivalence.
For every integer $m\ge1$ satisfying $m(m+1)<q-q/p$, the class
$\mathcal C_m$ properly contains the class of PN functions and includes
all their one-point modifications.

\Needspace{5\baselineskip}
\begin{corollary}[CCZ-invariant families and coincidence of CCZ- and EA-equivalence]
\label{cor:ccz-classes-near-PN}
Let $q=p^n>2$, where $p$ is prime.
\begin{enumerate}
\item[(i)] The family $\mathcal T$ satisfies
\[
 \mathcal C_{\CCZ}(\mathcal T)=\mathcal T
 =\mathcal C_{\EA}(\mathcal T).
\]
Equivalently, every function CCZ-equivalent to a function of maximal
algebraic degree also has maximal algebraic degree.
\item[(ii)] Suppose that $p$ is odd and $m\ge0$ is an integer
satisfying $m(m+1)<q-q/p$.  Then, for every $F\in\mathcal C_m$,
\[
 \mathcal C_{\CCZ}(F)=\mathcal C_{\EA}(F).
\]
Moreover,
\[
 \mathcal C_{\CCZ}(\mathcal C_m)
 =\mathcal C_{\EA}(\mathcal C_m)=\mathcal C_m.
\]
Equivalently, if $F$ differs from a PN function on at most $m$ inputs,
then every function $H$ CCZ-equivalent to $F$ is EA-equivalent to $F$
and differs from some PN function on at most $m$ inputs.
\end{enumerate}
\end{corollary}

\begin{proof}
(i) Let $H\in\mathcal C_{\CCZ}(\mathcal T)$, and choose
$F\in\mathcal T$ with $F\sim_{\CCZ}H$.
Then there is an $\F_p$-affine permutation
$\mathcal A(v)=\mathcal L(v)+t$ sending $\mathcal G_F$ onto
$\mathcal G_H$, where $\mathcal L$ is invertible and $\F_p$-linear.
Since $\sum_{x\in\F_q}x=0$ for $q>2$ and $qt=0$, we have
\[
 \left(0,\sum_xH(x)\right)
 =\sum_{P\in\mathcal G_H}P
 =\mathcal L\!\left(\sum_{P\in\mathcal G_F}P\right)
 =\mathcal L\!\left(0,\sum_xF(x)\right).
\]
By \eqref{eq:maximal-degree-sum}, the vector on the right is nonzero,
because $F\in\mathcal T$ and $\mathcal L$ is invertible.
Thus $\sum_xH(x)\ne0$, and \eqref{eq:maximal-degree-sum} gives
$H\in\mathcal T$.
From $\mathcal T\subseteq\mathcal C_{\EA}(\mathcal T)
\subseteq\mathcal C_{\CCZ}(\mathcal T)\subseteq\mathcal T$,
we obtain the required equality.

(ii) Let $F\in\mathcal C_m$, choose a PN function $f$ with
$k=d(F,f)\le m$, and let $H:\F_q\to\F_q$ be any function satisfying
$F\sim_{\CCZ}H$.  Since
$k(k+1)\le m(m+1)<q-q/p$,
Lemma~\ref{lem:ccz-ea-onepoint} gives $F\sim_{\EA}H$.
As $H$ is arbitrary, this proves
$\mathcal C_{\CCZ}(F)\subseteq\mathcal C_{\EA}(F)$.
The reverse inclusion always holds because EA-equivalence implies
CCZ-equivalence.  Hence
$\mathcal C_{\CCZ}(F)=\mathcal C_{\EA}(F)$.

Let $H$ be EA-equivalent to $F$.  Then
\[
 H(M_1(x)+a_0)=M_2(F(x))+B(x)+b_0
 \qquad(x\in\F_q),
\]
where $M_1,M_2$ are invertible $\F_p$-linear maps, $B$ is
$\F_p$-linear, and $a_0,b_0\in\F_q$.
As in the proof of Theorem~\ref{thm:onepoint-ccz-rigidity}, define $h$ by
\[
 h(M_1(x)+a_0)=M_2(f(x))+B(x)+b_0
 \qquad(x\in\F_q).
\]
The function $h$ is PN because it is EA-equivalent to $f$.
Moreover, the bijectivity of $M_1$ and $M_2$ gives
$d(H,h)=d(F,f)\le m$.
Thus $H\in\mathcal C_m$, proving that $\mathcal C_m$ is invariant
under EA-equivalence.
Using $\mathcal C_{\CCZ}(F)=\mathcal C_{\EA}(F)$ for every
$F\in\mathcal C_m$, established above, and the EA-invariance of
$\mathcal C_m$, we obtain
\[
 \mathcal C_{\CCZ}(\mathcal C_m)
 =\bigcup_{F\in\mathcal C_m}\mathcal C_{\CCZ}(F)
 =\bigcup_{F\in\mathcal C_m}\mathcal C_{\EA}(F)
 =\mathcal C_{\EA}(\mathcal C_m)=\mathcal C_m.
\]
\end{proof}

In Corollary~\ref{cor:ccz-classes-near-PN}(ii), the case $m=0$
recovers the result for planar functions.
For every odd $q>3$, one has $2<q-q/p$, so the case $m=1$
applies to all one-point modifications of PN functions.

Corollary~\ref{cor:ccz-classes-near-PN}(i) applies in particular to
functions over $\F_{2^n}$ for $n\ge2$: having algebraic degree $n$
is CCZ-invariant.

An analogue of Corollary~\ref{cor:ccz-classes-near-PN}(i) concerns
functions with the maximum possible number of bent components.  Although the number of bent components is not CCZ-invariant
in general, Mesnager et al.~\cite[Theorem~1]{MesnagerZhangTangZhou2019}
proved that, for even $n$,
having the maximum possible number $2^n-2^{n/2}$ of bent components
is CCZ-invariant for functions $F:\F_{2^n}\to\F_{2^n}$.

\paragraph{One-point modifications and maximal algebraic degree.}
\begin{corollary}[Maximal algebraic degree]
\label{cor:onepoint-maximal-degree}
Let $q=p^n>3$, where $p$ is odd, and let $f:\F_q\to\F_q$ be PN.
If $F:\F_q\to\F_q$ satisfies $d(F,f)=1$, then
\[
 \deg_{\mathrm{alg}}F=n(p-1).
\]
Every $H\in\mathcal C_{\CCZ}(F)$ has the same algebraic degree.
In particular, every $G_{\lambda,M}$ in Theorem~\ref{thm:even-PN-onepoint}
has maximal algebraic degree $n(p-1)$.
\end{corollary}

\begin{proof}
By the algebraic-degree bound for PN functions
\cite[Theorem~1.2]{BeierleBeyne2025}, we have
\[
 \deg_{\mathrm{alg}}f\le\frac{n(p-1)}2+1<n(p-1),
\]
where the strict inequality follows from $q>3$.
Hence \eqref{eq:maximal-degree-sum} gives $\sum_{x\in\F_q}f(x)=0$.
Let $a$ be the unique input at which $F(a)\ne f(a)$.  Then
\[
 \sum_x F(x)
 =F(a)+\sum_{x\ne a}F(x)
 =F(a)-f(a)+\sum_x f(x)
 =F(a)-f(a)\ne0,
\]
so \eqref{eq:maximal-degree-sum} gives
$\deg_{\mathrm{alg}}F=n(p-1)$.

The assertion for every $H\in\mathcal C_{\CCZ}(F)$ follows from
Corollary~\ref{cor:ccz-classes-near-PN}(i).
Finally, $g+M$ is PN and $G_{\lambda,M}$ differs from it only at zero,
by $-\lambda\ne0$.
In this case $\sum_xG_{\lambda,M}(x)=-\lambda$, so its
$x^{q-1}$-coefficient is exactly $\lambda$.
\end{proof}

For every odd prime $p$ and every $q=p^n>3$, our construction therefore
yields APN functions with boomerang uniformity one and algebraic
degree $n(p-1)$, the largest possible algebraic degree of a function
on $\F_q$.  The degree conclusion also holds for the cases with
boomerang uniformity two.

\Needspace{5\baselineskip}
This contrasts with the binary case:
Budaghyan et al.~\cite[Conjecture~1]{BudaghyanEtAl2018} conjectured that every APN
function on $\F_{2^n}$, $n\ge3$, has algebraic degree at most $n-1$.
The connection with one-point modifications is developed further in
\cite{BudaghyanCarletHellesethKaleyski2020}, and the degree conjecture
remains open in the recent work of Mihaila and
Thornburgh~\cite{MihailaThornburgh2026}.

We now apply Theorem~\ref{thm:onepoint-ccz-rigidity} to $G_{\lambda,M}$
and determine when varying $\lambda$, with $g$ and $M$ fixed, produces
CCZ-equivalent functions.
Let $g:\F_q\to\F_q$ be an even PN function satisfying $g(0)=0$,
where $q=p^n>3$ and $p$ is odd.  Define the set $\Lambda_g$ of
permitted values of $\lambda$ and the group $\mathcal M_g$ by
\begin{equation}\label{eq:parameter-orbit-definitions}
\begin{aligned}
 \Lambda_g&:=\{\lambda\in\F_q^*:-\lambda\notin g(\F_q)\},\\
 \mathcal M_g&:=\{M_2\in\operatorname{GL}_{\F_p}(\F_q):
    g\circ M_1=M_2\circ g\text{ for some }
    M_1\in\operatorname{GL}_{\F_p}(\F_q)\}.
\end{aligned}
\end{equation}
Here $\operatorname{GL}_{\F_p}(\F_q)$ denotes the group of
$\F_p$-linear permutations of $\F_q$.  By
Lemma~\ref{lem:character-preimage-counts}(i), we have
\[
 \Lambda_g=
 \begin{cases}
  g(\F_q^*),&\chi(-1)=-1,\\[2pt]
  \F_q^*\setminus g(\F_q^*),&\chi(-1)=1.
 \end{cases}
\]
Composition and inversion show that
$\mathcal M_g$ is a group.  Every $M_2\in\mathcal M_g$ maps
$g(\F_q)$ onto itself and therefore also maps $\Lambda_g$ onto itself.
The group $\mathcal M_g$ acts on $\Lambda_g$ by
$\lambda\mapsto M_2(\lambda)$.  For $\lambda\in\Lambda_g$, its orbit is
\[
 \mathcal M_g\cdot\lambda
 :=\{M_2(\lambda):M_2\in\mathcal M_g\}\subseteq\Lambda_g.
\]

\Needspace{12\baselineskip}
\begin{corollary}
\label{cor:parameter-orbits}
With the notation in \eqref{eq:parameter-orbit-definitions}, fix $g$
and an $\F_p$-linearized polynomial $M$, and consider the family
\[
 \{G_{\lambda,M}:\lambda\in\Lambda_g\}.
\]
\begin{enumerate}
\item[(i)] This family consists of $(q-1)/2$ APN functions.  For distinct
$\lambda,\mu\in\Lambda_g$, replacing only the value
$G_{\lambda,M}(0)=-\lambda$ by $-\mu$ produces the APN function
$G_{\mu,M}$.  Thus any two distinct members have Hamming distance
one.  If $M$ is a permutation, every member has boomerang uniformity
one.
\item[(ii)] For any two members $G_{\lambda,M}$ and $G_{\mu,M}$ of this family,
\[
 G_{\lambda,M}\sim_{\CCZ}G_{\mu,M}
 \quad\Longleftrightarrow\quad
 \mu\in\mathcal M_g\cdot\lambda.
\]
Equivalently, the two functions are CCZ-inequivalent precisely when
$\lambda$ and $\mu$ belong to different $\mathcal M_g$-orbits.
The CCZ-equivalence classes within this family therefore correspond
bijectively to the $\mathcal M_g$-orbits on $\Lambda_g$.
\end{enumerate}
\end{corollary}

\begin{proof}
(i) Lemma~\ref{lem:even-PN-preimages} gives $|\Lambda_g|=(q-1)/2$,
and Theorem~\ref{thm:even-PN-onepoint} gives the asserted
uniformities.  For $\lambda,\mu\in\Lambda_g$,
\[
 G_{\mu,M}(x)-G_{\lambda,M}(x)
 =\begin{cases}
   \lambda-\mu,&x=0,\\
   0,&x\ne0.
  \end{cases}
\]
Thus distinct values $\lambda,\mu\in\Lambda_g$ give functions at
Hamming distance one.

(ii) Taking $h=g$ and $N=M$ in
Theorem~\ref{thm:onepoint-ccz-rigidity} gives
\[
 G_{\lambda,M}\sim_{\CCZ}G_{\mu,M}
 \quad\Longleftrightarrow\quad
 \mu=M_2(\lambda)\text{ for some }M_2\in\mathcal M_g,
\]
which proves the equivalence criterion and the stated correspondence.
\end{proof}

\begin{remark}
\label{rem:distance-one-orbits}
Part~(i) contrasts with the binary single-point modification
conjecture~\cite[Conjecture~2]{BudaghyanEtAl2018}: for $n\ge3$,
changing exactly one value of an APN function on $\F_{2^n}$ is
conjectured to destroy the APN property.
As noted in the Introduction, the existence of
odd-characteristic APN pairs at Hamming distance one is already
implicit in Zha and Hu's construction~\cite{ZhaHu2013} based on
the PN function $g(x)=ux^2$, where $u\ne0$.
After output scaling and translation, their APN family corresponds
to $g(x)=x^2$ and $M=0$ in our construction, and hence has boomerang
uniformity two by Theorem~\ref{thm:even-PN-onepoint}.
In contrast, part~(i) applies to every even PN function $g$
satisfying $g(0)=0$ and yields pairs with boomerang uniformity one
whenever $M$ is a linearized permutation.

More generally, the criterion in part~(ii) extends to arbitrary
$\F_p$-linearized polynomials $M,N$: by
Theorem~\ref{thm:onepoint-ccz-rigidity}, for $\lambda,\mu\in\Lambda_g$,
one has $G_{\lambda,M}\sim_{\CCZ}G_{\mu,N}$ if and only if
$\mu=M_2(\lambda)$ for some $M_2\in\mathcal M_g$.
Thus the criterion for CCZ-equivalence is independent of the choices
of $M$ and $N$.

Corollary~\ref{cor:monomial-parameter-equivalence} and
Proposition~\ref{prop:DY-distance-one} below illustrate the transitive
and nontransitive cases, respectively, of the action of
$\mathcal M_g$ on $\Lambda_g$.
\end{remark}

\begin{corollary}
\label{cor:monomial-parameter-equivalence}
Suppose that $g(x)=x^d$ is an even PN function on $\F_q$.
For every $\lambda,\mu\in\Lambda_g$ and all $\F_p$-linearized
polynomials $M,N$, the functions $G_{\lambda,M}$ and
$G_{\mu,N}$ are EA-equivalent, and hence CCZ-equivalent.
More precisely, there exists $a\in\F_q^*$ such that
$a^d=\mu/\lambda$, and
\begin{equation}\label{eq:monomial-parameter-EA}
 G_{\mu,N}(ax)
 =a^dG_{\lambda,M}(x)+N(ax)-a^dM(x)
 \qquad(x\in\F_q).
\end{equation}
In particular, for distinct $\lambda,\mu\in\Lambda_g$, the APN functions
$G_{\lambda,M}$ and $G_{\mu,M}$ are EA-equivalent, and each is a
one-point modification of the other.  If $M$ is a permutation, both
functions have boomerang uniformity one.
\end{corollary}

\begin{proof}
By Lemma~\ref{lem:even-PN-preimages}, the map $x\mapsto x^d$
is two-to-one on $\F_q^*$, so $\gcd(d,q-1)=2$ and
$g(\F_q^*)$ is the set of nonzero squares.  Both $-\lambda$ and
$-\mu$ are nonsquares.  Thus $\mu/\lambda$ is a square, and there
exists $a\ne0$ with $a^d=\mu/\lambda$.
For $x\ne0$, identity~\eqref{eq:monomial-parameter-EA} follows
from $g(ax)=a^dg(x)$; at $x=0$, both sides equal $-\mu$.
The remaining assertions follow from Corollary~\ref{cor:parameter-orbits}(i).
\end{proof}

This applies to all Coulter--Matthews and Gold PN functions used in
Sections~\ref{sec:coulter-matthews} and~\ref{sec:DO-examples}.
For a nonmonomial PN function $g$, as shown below, different choices of
$\lambda\in\Lambda_g$ can give CCZ-inequivalent functions
$G_{\lambda,M}$, even when the same linearized permutation $M$
is used throughout.

\begin{proposition}[CCZ-inequivalent Ding--Yuan modifications at Hamming distance one]
\label{prop:DY-distance-one}
Let $\F_{3^5}=\F_3[t]/(t^5-t-1)$ and
$g(x)=x^{10}-x^6-x^2$.  Then $-1,1+t\in\Lambda_g$.
For every $\F_3$-linearized polynomial $M$, the two APN functions
$G_{-1,M}$ and $G_{1+t,M}$ are CCZ-inequivalent, and each is a
one-point modification of the other.  If $M$ is a permutation, both
functions have boomerang uniformity one.
\end{proposition}

\begin{proof}
The function $g$ is a Ding--Yuan PN function~\cite{DingYuan2006}.
Put $s=t^4+t^3-t^2$.  Reduction using $t^5=t+1$ gives
$g(1)=-1$ and $g(s)=1+t$.
Since $\chi(-1)=-1$, Lemma~\ref{lem:character-preimage-counts}(i)
gives $\Lambda_g=g(\F_{3^5})\setminus\{0\}$.  Thus
$-1,1+t\in\Lambda_g$.
By Corollary~\ref{cor:parameter-orbits}(i), $G_{-1,M}$ and
$G_{1+t,M}$ are APN functions, each a one-point modification of the
other.  If $M$ is a permutation, both functions have boomerang
uniformity one.

We further show that this one-point modification produces
CCZ-inequivalent APN functions.
Write $L_a=L_{a,g}$ for the linearized derivative defined in
Section~\ref{sec:preliminaries}.
For $a,x\in\F_{3^5}$, we have
\[
 L_a(x)=g(x+a)-g(x)-g(a)
       =(a^9+a)x+a^3x^3+ax^9.
\]
Thus $L_a$ is $\F_3$-linear, and it is a permutation whenever
$a\ne0$, because $g$ is PN.
For an $\F_3$-linear transformation $U:\F_{3^5}\to\F_{3^5}$, let
$\operatorname{tr}(U)$ denote its trace, which is independent of
the chosen $\F_3$-basis.
For $a\ne0$, define
\begin{equation}\label{eq:DY-trace-form}
 T_a(x,y):=\operatorname{tr}\bigl(
 L_a^{-1}\circ L_x\circ L_a^{-1}\circ L_y\bigr).
\end{equation}
The form $T_a$ is $\F_3$-bilinear because $x\mapsto L_x$ is
$\F_3$-linear, and it is symmetric because
$\operatorname{tr}(UV)=\operatorname{tr}(VU)$.
Moreover, $L_{-a}=-L_a$ gives $T_{-a}=T_a$.

Suppose that $G_{-1,M}\sim_{\CCZ}G_{1+t,M}$.
Then Theorem~\ref{thm:onepoint-ccz-rigidity} supplies invertible
$\F_3$-linear maps $M_1,M_2$ satisfying
\[
 g(M_1(x))=M_2(g(x))\quad(x\in\F_{3^5}),\qquad M_2(-1)=1+t.
\]
Linearizing $g\circ M_1=M_2\circ g$ gives
$L_{M_1(x)}\circ M_1=M_2\circ L_x$.
Hence, for $a\ne0$,
\[
 L_{M_1(a)}^{-1}\circ L_{M_1(x)}
 =M_1\circ(L_a^{-1}\circ L_x)\circ M_1^{-1}.
\]
Applying this identity for $x$ and $y$, and using invariance of
matrix trace under conjugation, we obtain
\begin{equation}\label{eq:DY-trace-congruence}
 T_{M_1(a)}(M_1(x),M_1(y))=T_a(x,y).
\end{equation}
Since $g(M_1(1))=M_2(g(1))=1+t=g(s)$,
Lemma~\ref{lem:even-PN-preimages} gives $M_1(1)=\pm s$.
Taking $a=1$ in \eqref{eq:DY-trace-congruence} and using $T_{-s}=T_s$,
we obtain
\[
                         T_s(M_1(x),M_1(y))=T_1(x,y).
\]

With respect to the $\F_3$-basis $1,t,t^2,t^3,t^4$, the Gram matrix
of a bilinear form $T$ is
$[T]=\bigl(T(t^i,t^j)\bigr)_{0\le i,j\le4}$.
Applying \eqref{eq:DY-trace-form}, we obtain
\[
 [T_1]=
 \begin{pmatrix}
 2&0&0&0&1\\
 0&2&2&1&2\\
 0&2&1&1&1\\
 0&1&1&2&0\\
 1&2&1&0&1
 \end{pmatrix},\qquad
 [T_s]=
 \begin{pmatrix}
 2&1&2&0&1\\
 1&2&1&2&0\\
 2&1&0&2&1\\
 0&2&2&0&0\\
 1&0&1&0&2
 \end{pmatrix}.
\]
These matrices follow by expanding
$L_z(t^j)=(z^9+z)t^j+z^3t^{3j}+zt^{9j}$ in this basis,
using $t^5=t+1$, and applying \eqref{eq:DY-trace-form}.
Their determinants satisfy
\[
               \det[T_1]=1,\qquad \det[T_s]=2
               \quad\text{in }\F_3.
\]
On the other hand, if $[M_1]$ is the matrix of $M_1$ in the same basis,
the identity $T_s(M_1(x),M_1(y))=T_1(x,y)$ gives
$[T_1]=[M_1]^{\mathsf T}[T_s][M_1]$.  Since
$\det[M_1]\in\F_3^*$, one has $(\det[M_1])^2=1$, which would imply
$\det[T_1]=\det[T_s]$, a contradiction.
This proves CCZ-inequivalence.
\end{proof}

In particular, taking $M(x)=x$ in
Proposition~\ref{prop:DY-distance-one} gives the explicit pair
\[
\begin{aligned}
 F(x)&=x^{10}-x^6-x^2+x-(x^{242}-1),\\
 H(x)&=x^{10}-x^6-x^2+x+(1+t)(x^{242}-1).
\end{aligned}
\]
Both functions have differential uniformity two and boomerang
uniformity one, and $F(x)=H(x)$ for every $x\ne0$, whereas
$F(0)=1$ and $H(0)=-1-t$.  Their CCZ-inequivalence shows that
changing one function value can preserve both the differential and
boomerang uniformities while changing the CCZ-equivalence class.

\section{Conclusion and further directions}\label{sec:conclusion}

We have shown that every even PN function $g:\F_q\to\F_q$ with
$g(0)=0$, where $q=p^n>3$ is odd, yields APN functions
$G_{\lambda,M}$ with boomerang uniformity at most two.
We have characterized precisely when this uniformity attains its
optimal value one; in particular, every linearized permutation $M$
gives this optimum.  The construction applies in every odd characteristic
and also allows even PN functions that are not DO, addressing two
directions raised in our previous work~\cite{KooKwonKoKim2026}.

We have determined complete differential and boomerang spectra.
The differential spectrum depends only on $q$ and, for $q>7$,
differs from those of all APN monomial families listed in
Table~\ref{tab:apn-power-functions} and every APN binomial in the
original Ness--Helleseth family or its extension by Zeng et al.
This proves CCZ-inequivalence to these functions.
We have obtained closed formulas for the boomerang spectra of
selected Coulter--Matthews examples and of our construction with
$M=L_c$ for arbitrary DO PN functions.  For Coulter--Matthews functions,
the number of values of $z(k,n)$ attained as $n$ varies is unbounded
as $k$ varies.  For the DO perturbations $M=L_c$, the number
of nonzero parameters $c$ yielding each boomerang spectrum depends
only on $q$.

Theorem~\ref{thm:onepoint-ccz-rigidity} characterizes CCZ-equivalence
within the construction and shows that any two linearly inequivalent
even PN functions over the same field yield CCZ-inequivalent APN functions.  Using nucleus orders, known inequivalence results, and
algebraic degrees to distinguish the underlying PN functions, we obtain
three pairwise
CCZ-inequivalent APN functions with boomerang uniformity one over
$\F_{p^{15\ell}}$ for every $\ell\ge1$, six over $\F_{p^{120r}}$
for every $r\ge1$, and unbounded numbers of such classes in
characteristic three.  Within each field, these pairwise CCZ-inequivalent
APN functions all have the same differential spectrum.

More generally, CCZ-equivalence and EA-equivalence coincide for
functions at Hamming distance at most $m$ from a PN function,
provided that $m(m+1)<q-q/p$.  Every nontrivial one-point modification of a PN
function has maximal algebraic degree $n(p-1)$.
For fixed $g$ and $M$, varying $\lambda$ in our construction also
produces APN functions at Hamming distance one.  Their CCZ classes are determined by the
$\mathcal M_g$-orbits on $\Lambda_g$, and our examples include both
CCZ-equivalent and CCZ-inequivalent pairs with boomerang uniformity one.

Two questions remain for further study: determining these parameter
orbits for other nonmonomial PN functions, especially Ding--Yuan
functions over larger odd-degree extensions, and evaluating
$z_{\lambda,M}$ for further choices of $g$ and $M$.
Their resolution would clarify how many CCZ classes arise from
a fixed PN function and how the boomerang spectra vary with the
perturbation.

\medskip
\noindent\textbf{Acknowledgments:} Soonhak Kwon was supported by Basic Science Research Program
through the National Research Foundation of Korea (NRF) funded by the Ministry of
Education (No.\ RS-2019-NR040081). Namhun Koo was supported by Basic Science Research
Program through the National Research Foundation of Korea (NRF) funded by the Ministry
of Education (No.\ RS-2026-25575984).

\end{document}